\documentstyle[12pt,subeqnarray,psfig]{article}
\begin{document}
%
%
%
\newenvironment{lefteqnarray}{\arraycolsep=0pt\begin{eqnarray}}
{\end{eqnarray}\protect\aftergroup\ignorespaces}
\newenvironment{lefteqnarray*}{\arraycolsep=0pt\begin{eqnarray*}}
{\end{eqnarray*}\protect\aftergroup\ignorespaces}
\newenvironment{leftsubeqnarray}{\arraycolsep=0pt\begin{subeqnarray}}
{\end{subeqnarray}\protect\aftergroup\ignorespaces}
\newcommand{\appleq}{\stackrel{<}{\sim}}
\newcommand{\appgeq}{\stackrel{>}{\sim}}
\newcommand{\arcsinh}{\mathop{\rm arcsinh}\nolimits}
\newcommand{\arctg}{\mathop{\rm arctg}\nolimits}
\newcommand{\diff}{{\rm\,d}}
\newcommand{\displayfrac}[2]{\frac{\displaystyle #1}{\displaystyle #2}}
\newcommand{\Erfc}{\mathop{\rm Erfc}\nolimits}
\newcommand{\Int}{\mathop{\rm Int}\nolimits}
\newcommand{\Nint}{\mathop{\rm Nint}\nolimits}
\newcommand{\pprime}{{\prime\prime}}

\title{A necessary condition for best fitting analytical to simulated 
       density profiles 
       in dark matter haloes}
\author{{R. Caimmi\footnote{
{\it Astronomy Department, Padua Univ., Vicolo Osservatorio 2,
I-35122 Padova, Italy}
email: caimmi@pd.astro.it}
\phantom{agga}}}
%
%
%

\maketitle
\begin{quotation}
\section*{}
\begin{Large}
\begin{center}

Abstract
\end{center}
\end{Large}
\begin{small}


Analytical and geometrical properties
of generalized power-law (GPL) density profiles
are reviewed, and special effort is devoted to
the special cases where GPL density profiles
reduce to (i) a double power-law (DPL), and (ii)
a single power-law (SPL).
Then GPL density profiles are compared with
simulated dark haloes (SDH) density profiles,
and nonlinear least-squares fits are prescribed,
involving five parameters (a scaling radius,
$r_0$, a scaling density, $\rho_0$, and three
exponents, $\alpha$, $\beta$, $\gamma$), which
specify the fitting profile (RFSM5 method).
More specifically, the validity of a necessary
condition for the occurrence of an extremal
point, is related to the existence of an
intersection between three surfaces in a
3-dimension space.   Using the algorithm
makes also establish that the extremal point
is a fiducial minimum, while the explicit
calculation of the Hessian determinant is
avoided to gain in simplicity.   In absence
of a rigorous proof, the fiducial minimum
can be considered as nothing but a fiducial 
absolute minimum.   An application is
made to a sample of 17 SDHs on the scale of
cluster of galaxies, within a flat $\Lambda$CDM
cosmological model (Rasia et al. 2004).   In dealing
with the averaged SDH density profile (ADP), a virial
radius, $r_{vir}$, equal to the mean over the whole
sample, is assigned, which allows the calculation
of the remaining parameters.   The following
results are found.   (i) A necessary condition
for the occurrence of an extremal point is
satisfied for 8 sample haloes, and is not for
the remaining 9 together with ADP.    In the
former alternative, an extremal minimum point
(EMP) may safely exist.   In the latter alternative,
the occurrence of an EMP cannot be excluded,
but only a non extremal minimum can be determined.
(ii) The occurrence of an EMP implies a sum of
square residuals which is systematically lower
than its counterpart deduced by use of numerical
RFSM5 methods.   Accordingly, EMPs may safely be
thought of as absolute minima.   With regard to
sample haloes where no EMP is detected, the above
result maintains in 4 cases (including ADP),
while the contrary holds for the remaining 5
cases.   (iii) The best fit (with no EMP detected)
is provided by DPL density profiles for 3 sample
haloes.   In addition, DPL density profiles make
a rough, but viable approximation in fitting SDH
density profiles.   The contrary holds for SPL
density profiles.   No evident correlation is 
found between SDH dynamical state (relaxed or 
merging) and asymptotic inner slope of the 
logarithmic density profile or (for SDH comparable
virial masses) scaled radius.   Mean values and 
standard deviations of some parameters are calculated
and, in particular, the decimal logarithm of the
scaled radius, $\xi_{vir}$,
reads $<\log\xi_{vir}>\approx0.59$ and $\sigma_{s
\log\xi_{vir}}\,\approx0.59$, the standard deviation
exceeding by a factor 3.3-3.4 its counterpart
evaluated in an earlier attempt using NFW (Navarro
et al. 1995, 1996, 1997) density profiles (Bullock
et al. 2001).
If a large dispersion still maintains for richer
samples, in dealing with analytical RSFM5 methods,
a low dispersion found in $N$-body simulations seems
to be an artefact,
due to the assumption of NFW (or any equivalent
choice) density profile.
It provides additional support to the idea, that
NFW density profiles may be considered as a
convenient way to parametrize SDH density
profiles, without implying that it necessarily
produces the best possible fit (Bullock et al. 2001).
With regard to RFSM5 methods formulated
in the current paper, the exponents of both the
best fitting GPL density profile to ADP, $(\alpha,
\beta,\gamma)\approx(0.3,4.5,1.5)$, and related
averages calculated over the whole halo sample,
$(\overline{\alpha},\overline{\beta},\overline
{\gamma})\approx(3.0,3.9,1.3)$, are far from
their NFW counterparts, $(\alpha,\beta,\gamma)
=(1,3,1)$.   The last result, together with
the large value of the standard deviation,
$\sigma_{s\log\xi_{vir}}$, is interpreted as
due to a certain degree of degeneracy in
fitting GPL to SDH density profiles.
If it is a real feature of the problem, or it 
could be reduced by the next generation of
high-resolution simulations, still remains
an open question.   Values of asymptotic inner slope
of fitting logarithmic density profiles, are
consistent with results from recent high-resolution
simulations (Diemand et al. 2004; Reed et al. 2004).

\noindent
{\it keywords - cosmology: dark matter -- galaxies:
clusters:general.}



\end{small}
\end{quotation}


\section{Introduction}\label{intro}
Recent observations from anisotropies in the cosmic
microwave background, large-scale structure surveys,
Hubble parameter determinations, and Type Ia
supernovae results, allow narrow ranges for the
values of cosmological quantities (e.g., Sievers et al.
2003; Rubi$\tilde{{\rm n}}$o-Martin et al. 2003; Spergel
et al. 2003).
The related ($\Lambda$CDM) cosmological model is 
consistent with a bottom-up picture (hierarchical
clustering) of dark matter haloes, where smaller
systems formed first from initial density 
perturbations and then merged with each other to
become larger systems, or were tidally disrupted
and accreted from previously formed larger systems.

A two-parameter functional form for the
halo profile, where a scaled density depends on a
scaled radius and three exponents are specified,
has been proposed by several authors (e.g., Navarro
et al. 1995, 1996, 1997, hereafter quoted together
as NFW; Moore et al. 1998, 1999; Rasia et al. 2004).
Generally speaking, it may be considered as a special
case of the 5-parameter family (Hernquist 1990):
\begin{equation}
\label{eq:GPL}
\rho\left(\frac r{r_0}\right)=\frac{\rho_0}{(r/r_0)^\gamma
[1+(r/r_0)^\alpha]^\chi}~~;\quad\chi=\frac{\beta-\gamma}
\alpha~~;
\end{equation}
where $\rho_0$ and $r_0$ are a scaling density and a
scaling radius, respectively, and all the exponents
are positive in the case under discussion.

The density profile, expressed by Eq.\,(\ref{eq:GPL}),
reduces to a power-law both towards the centre, $r\to0$,
and towards infinite, $r\to+\infty$, where the exponent
equals $-\gamma$ and $-\beta$, respectively.
It may be conceived as a generalized power-law and, in the
following, it shall be quoted as GPL density profile.
On the other hand, matter distribution within a
simulated dark matter halo, in the following, shall be
quoted as SDH density profile.

For fixed exponents, one among the two remaining
free parameters, the scaling density and the scaling
radius, may be related to the mass and the radius of
the virialized region.   For further details see e.g.,
NFW; Bullock et al. (2001); Rasia et al. (2004);
Caimmi et al. (2004; hereafter quoted as CMV).

Strictly speaking, GPL density profiles cannot be
considered as universal, in the sense that a fixed
choice of exponents, $(\alpha,\beta,\gamma)$, does
not provide the best fit to an assigned set of SDH
density profiles (e.g., Bullock et al. 2001; CMV).
A weaker conception of universal, GPL density
profiles, might be related to the insensitivity of
the shape to halo masses, initial density fluctuation
spectra, or cosmological parameters (NFW; Hess et al.
1999; Klypin et al. 2001; Bullock et al. 2001;
Fukushige \& Makino 2001, 2003; Zhao et al. 2003).

However, no general consensus still exists on
this point.   In particular, some authors found
an inner slope which depends on the power spectrum
of the initial density perturbation (Syer \&
White 1998), or on the mass (Jing \& Suto 2000;
Hiotelis 2003; Ricotti 2003).   Further support
to the above results is provided by recent,
high-resolution, $N$-body simulations (Fukushige
et al. 2004; Navarro et al. 2004).  In addition, a
certain degree of degeneracy occurs,
with regard to the exponents, $(\alpha,\beta,
\gamma)$, in fitting various GPL to SDH density
profiles, in the whole range of resolved scales 
(e.g., Klypin et al. 2001).   For a more
detailed discussion, see CMV.

According to the above remarks, GPL density
profiles may be considered as a convenient
way to parametrize SDH density profiles,
without implying that it necessarily produces
the best possible fit (Bullock et al. 2001).
The search of the best fitting GPL to (one
or more) SDH density profiles, makes the aim
of the current paper.   To this respect,
different fitting criteria exist, such as
minimizing the maximum fractional deviations
of the fit, $\max\vert\log(\rho_{GPL}/\rho_h)
-\log(\rho_{SDH}/\rho_h)\vert$ (e.g., Klypin
et al. 2001); the sum of the squares of absolute%
\footnote{The term ``absolute'' here has not
to be intended as ``absolute value'', but as 
opposite to ``relative''.   More precisely,
$y_i-y(x_i)$ is an absolute residual, while
$[y_i-y(x_i)]/y(x_i)$ is the corresponding
relative residual.}%
logarithmic
residuals, $\chi^2=\sum[\log(\rho_{GPL}/\rho_h)-
\log(\rho_{SDH}/\rho_h)]^2$ (e.g., Bullock et al.
2001); the sum of squares of relative residuals,
$\sum[(\rho_{SDH}/\rho_h-\rho_{GPL}/\rho_h)/(\rho_
{GPL}/\rho_h)]^2$ (e.g., Fukushige \& Makino 2003,
2004), where $\rho_h$ is a normalization value.
For a more detailed discussion see e.g., Tasitsiomi
et al. (2004).

The hierarchical collapse of dark matter into
virialized haloes is likely to have played a
key role in the formation of large-scale objects,
such as galaxies and clusters of galaxies.   The
halo density distribution has a direct dynamical
role in determining the observable parameters of the
baryonic subsystems.   Then further investigation
even on the best fitting GPL to SDH density profile,
appears to be important.

To this aim, a nonlinear least-squares method
is used in the current paper.   The related,
general procedure, shall be quoted in the
following as RFSM5 (Residuals Functions Sum
Minimization within a 5-dimension hyperspace)
method.   The radius of the virialized region
is taken from computer outputs, for selected
cosmological parameters, which allows to deduce
the mass of the virialized region, and the
remaining quantities of interest.   Further
details may
be found in an earlier attempt (CMV), where
fiducial minima of the sum of absolute values
and squares of residuals are calculated for
assigned SDH density profiles, using a
5-dimension hypergrid within a 5-dimension
hyperspace.

An analytical procedure, instead of the
above mentioned numerical, shall be exploited
in the current paper.   The problem may be
formulated from both an analytical and
geometrical point of view.   The sum of
square residuals is a function of 5 variables
in a 6-dimension hyperspace, $w=F(r_0,\rho_0,
\alpha,\beta,\gamma)$ which, in turn, represents
a 5-dimension hypersurface.  The extremal point
of absolute minimum defines the best fitting
GPL to SDH density profile under consideration.
If more than one extremal points of absolute
minimum exist, then degeneracy occurs.   For
further details, see CMV.

According to a theorem of mathematical analysis,
a necessary condition for the occurrence of an
extremal point of minimum, is a null (or undefined)
value of all the first partial derivatives of
an assigned function, evaluated at that point.
A study of the Hessian determinant enables to
establish additional features (minimum, maximum,
hyperflexion, hypersaddle).   Finally, one has
to ensure that the minimum is an absolute minimum.
For further details on the analytic procedure
see e.g., Smirnov (1969, Chap.\,V, \S\S\,2.2-2.6).

In fitting GPL to SDH density profiles, our
attention shall be restricted to the sum of
square residuals, as absolute values exhibit
a discontinuity of the first derivative for
null arguments, which makes a solution of the
problem much more difficult.   An algorithm
will be formulated and used, for searching
points where the above mentioned necessary
condition for the occurrence of an extremal
minimum, is satisfied.   Then it will be
inferred, by comparison with numerical
results found in an earlier attempt (CMV),
that the related point is a fiducial minimum.
On the other hand, no proof will be produced,
that the point under discussion is an absolute
minimum.    This, and other improvements, are
left to mathematicians.

The fit of GPL to SDH density profiles, which
has deeply been investigated in CMV, is
reviewed in section \ref{fGS}.   The analytical
RFSM5 method used for fulfilling the necessary
condition for the occurrence of an extremal
point of minimum, is presented in section \ref
{RFSM5} and derived in Appendix A.   The subject of
section \ref{ara} is an application to a sample
of 17 SDHs and the related mean density profile,
on the scale of clusters of galaxies, taken from
Rasia et al. (2004), and already used in CMV.
Then the results are discussed.
Some concluding remarks are drawn in section 
\ref{conc}.

\section{Fitting GPL to SDH density profiles}
\label{fGS}

A detailed investigation on GPL and SDH density
profiles, together with their comparison, has
been performed in an earlier attempt (CMV),
to which an interested
reader is addressed.   Only what is relevant
for the current paper, or has not been mentioned
in CMV, shall be reported here.

\subsection{GPL density profiles}\label{GPL}

Plotting GPL density profiles on a logarithmic
plane, $({\sf O}~\log\xi~\log f)$, necessarily
implies use of dimensionless coordinates, 
defined as:
\begin{equation}
\label{eq:csif}
\xi=\frac r{r_0}~~;\qquad f(\xi)=\frac\rho{\rho_0}~~;
\end{equation}
where $\xi$ and $f$ can be conceived as a scaled
radius and a scaled density, respectively.
Accordingly, Eq.\,(\ref{eq:GPL}) reduces to:
\begin{equation}
\label{eq:SGPL}
f(\xi)=\frac1{\xi^\gamma(1+\xi^\alpha)^\chi}~~;\qquad 
\chi=\frac{\beta-\gamma}\alpha~~;
\end{equation}
independent of the scaling parameters.

As Eqs.\,(\ref{eq:csif}) and (\ref{eq:SGPL})
imply null density at infinite radius, the
mass distribution has necessarily to be 
ended at an assigned isopycnic surface,
which defines a truncation radius, $R$.
The mass within the truncation isopycnic
surface is (Caimmi \& Marmo 2003; CMV):
\begin{equation}
\label{eq:Mt}
M=M(R)=\frac{4\pi}3r_0^3\rho_0\nu_{mas}=M_0\nu_{mas}~~;
\end{equation}
where $M_0$ is a scaling mass and the profile factor,
$\nu_{mas}$, has the explicit expression:
\begin{equation}
\label{eq:num}
\nu_{mas}=3\int_0^\Xi f(\xi)\xi^2\diff\xi~~;
\end{equation}
and the integration is carried up to:
\begin{equation}
\label{eq:Csi}
\Xi=\frac R{r_0}~~;
\end{equation}
which may be conceived as a scaled, truncation
radius.

Two limiting cases of Eq.\,(\ref{eq:SGPL}) can
be relevant to the aim of the current paper.
The first one occurs as $\alpha\to+\infty$,
which implies $\chi\to0$ provided $\beta$ and
$\gamma$ remain finite, and:
\begin{lefteqnarray*}
&& (1+\xi^\alpha)^\chi\sim1~~;\qquad\xi\le1~~; \\
&& (1+\xi^\alpha)^\chi\sim\xi^{\alpha\chi}\sim
\xi^{\beta-\gamma}~~;\qquad\xi>1~~;
\end{lefteqnarray*}
accordingly, Eq.\,(\ref{eq:SGPL}) reduces to:
\begin{equation}
\label{eq:liai}
\lim_{\alpha\to+\infty}f(\xi)=\cases{\xi^{-\gamma}
~~;\qquad\xi\le1~~;\qquad r\le r_0~~; & \cr
\xi^{-\beta}~~;\qquad\xi>1~~;\qquad r>r_0~~; \cr}
\end{equation}
which is a double power-law with exponents
equal to $-\gamma$ and $-\beta$.

The second limiting case of Eq.\,(\ref{eq:SGPL})
occurs as $\beta\to+\infty$, which implies $\chi
\to+\infty$ provided $\alpha$ and $\gamma$ remain
finite.   If, in addition, $r_0\to+\infty$ i.e.
$\Xi\to0$, and:
\begin{lefteqnarray*}
&& (1+\xi^\alpha)^\chi\sim\frac1{C_\xi}~~;\qquad
\rho_0\sim C_\rho\Xi^\gamma~~;
\end{lefteqnarray*}
where $C_\xi$ and $C_\rho$ are constants, then
Eq.\,(\ref{eq:SGPL}) reduces to:
\begin{equation}
\label{eq:libi}
\lim_{\beta\to+\infty}f(\xi)=\frac{C_\xi}
{\xi^\gamma}~~;
\end{equation}
A similar result is also obtained
as $\alpha\to0$ provided $\chi$ remains constant,
which implies $\beta\to\gamma$ and:
\begin{lefteqnarray*}
&& (1+\xi^\alpha)^\chi\sim2^\chi~~;
\end{lefteqnarray*}
accordingly, Eq.\,(\ref{eq:SGPL}) reduces to:
\begin{equation}
\label{eq:lia0}
\lim_{\alpha\to0}f(\xi)=2^{-\chi}\xi^{-\gamma}~~;
\end{equation}
which is a power-law with exponent equal to
$-\gamma$.

The logarithmic GPL density profile, deduced
from Eq.\,(\ref{eq:SGPL}), is:
\begin{equation}
\label{eq:logf}
\log f=-\gamma\log\xi-\chi\log(1+\xi^\alpha)~~;
\end{equation}
and the maximum variation in slope occurs
at $\log\xi=\log(r/r_0)=0$ i.e. $r=r_0$,
which discloses the geometrical meaning of
the scaling radius in GPL density profiles.
In the logarithmic plane $({\sf O}~\log\xi
~\log f)$, Eq.\,(\ref{eq:logf}) represents
a curve where two asymptotes exist:
\begin{leftsubeqnarray}
\slabel{eq:as0}
&& \log f=-\gamma\log\xi~~;\qquad\xi\ll1~~; \\
\slabel{eq:asi}
&& \log f=-\beta\log\xi~~;\qquad\xi\gg1~~;
\label{seq:as}
\end{leftsubeqnarray}
which meet at the origin.   For further
details, see CMV.

\subsection{SDH density profiles}\label{SDH}

Simulated haloes are characterized by a
``virial'' parameter, either the virial
mass, $M_{vir}$, or the virial radius,
$r_{vir}$, defined such that the mean
density inside the virial radius is
$\Delta_{vir}$ times the mean matter
universal density, $\rho_h=\rho_{crit}
\Omega_m$, at that redshift:
\begin{equation}
\label{eq:Mvir}
M_{vir}=\frac{4\pi}3\Delta_{vir}\rho_
{crit}\Omega_mr_{vir}^3~~;
\end{equation}
where $\rho_{crit}=3H^2/(8\pi G)$ is
the critical density for closure.
The critical overdensity at virialization,
$\Delta_{vir}$, is motivated by the
spherical collapse model: it is below
two hundreds for an Einstein-de Sitter
cosmology, and exceeds three hundreds
for a flat $\Lambda$CDM cosmology where
$\Omega_m\approx0.3$, at $z=0$ (e.g.,
Bullock et al. 2001).

Plotting SDH density profiles on a 
logarithmic plane, $({\sf O}\log\eta~
\log\psi)$, necessarily implies use of
dimensionless coordinates, defined as:
\begin{equation}
\label{eq:SSDH}
\eta=\frac r{r_{vir}}~~;\qquad\psi(\eta)=
\frac\rho{\rho_h}~~;\qquad\rho_h=\rho_
{crit}\Omega_m~~;
\end{equation}
where an upper and lower limit to the 
domain follow
from the definition of virial radius
and the occurrence of numerical artifacts,
respectively.   This is why (i) regions
outside the virial radius are still
falling in, and then macroscopic kinetic
energy has still to be converted into
peculiar energy (e.g., Cole \& Lacey 1996;
NFW), and (ii) two-body relaxation takes
place in the central regions of simulated
haloes, within about $0.01
r_{vir}$ (e.g., Bullock et al. 2001;
Fukushige \& Makino 2001, 2003, 2004;
Navarro et al. 2004).
Accordingly, SDH density profiles
within the range:
\begin{equation}
\label{eq:doeta}
-2<\log\eta<0~~;
\end{equation}
appear to be closely related to the
virialized region.

\subsection{Comparison between GPL and SDH
density profiles}\label{GSc}

A comparison between GPL and SDH density
profiles necessarily implies that the
truncation radius, and the mass enclosed
within the truncation isopycnic surface,
do coincide with the virial radius and
the virial mass, $R=r_{vir}$ and $M=M_
{vir}$, respectively.   Then the combination
of Eqs.\,(\ref{eq:Mt}), (\ref{eq:Mvir}), and 
(\ref{eq:SSDH}), yields:
\begin{lefteqnarray}
\label{eq:rho0h}
&& \frac{\rho_0}{\rho_h}=\frac{\Delta_{vir}\xi_{vir}^3}
{\nu_{mas}}~~; \\
\label{eq:csiv}
&& \xi_{vir}=\frac{r_{vir}}{r_0}~~;
\end{lefteqnarray}
where the scaled virial radius, $\xi_{vir}$,
may be taken as a definition of the
concentration (CMV).

Relating scaled GPL density profiles,
expressed by Eqs.\,(\ref{eq:csif}), to scaled
SDH density profiles, expressed by Eqs.\,(\ref
{eq:SSDH}), yields:
\begin{equation}
\label{eq:fpsi}
\xi=\xi_{vir}\eta~~;\qquad f=\frac{\rho_h}
{\rho_0}\psi~~;
\end{equation}
where $\xi_{vir}$ is defined by Eq.\,(\ref
{eq:csiv}).   Accordingly, a generic, scaled
GPL density profile, expressed by Eq.\,(\ref
{eq:SGPL}), takes the equivalent form:
\begin{equation}
\label{eq:PGPL}
\psi(\eta)=\frac{\rho_0/\rho_h}{(\xi_{vir}\eta)^
\gamma\left[1+(\xi_{vir}\eta)^\alpha\right]^
\chi}~~;\qquad \chi=\frac{\beta-\gamma}\alpha~~;
\end{equation}
and the related, logarithmic GPL density
profile, is deduced by use of Eq.\,(\ref
{eq:rho0h}). The result is:
\begin{lefteqnarray}
\label{eq:lpsi}
&& \log\psi=\log\Delta_{vir}-\log\nu_{mas}+
3\log\xi_{vir} \nonumber \\
&& -\gamma\log\xi_{vir}-\gamma\log\eta-\chi
\log\left[1+(\xi_{vir}\eta)^\alpha\right]~~;
\end{lefteqnarray}
which depends on three exponents, $(\alpha,
\beta,\gamma)$, and two scaling parameters,
$(r_0,\rho_0)$.   On the other hand, the
scaled mass, $\nu_{mas}$, is defined by
Eq.\,(\ref{eq:num}); the virial radius,
$r_{vir}$, is known from the computer run;
and the critical overdensity, $\Delta_{vir}$,
is determined by the cosmological model.

In the logarithmic plane, $({\sf O}\log\eta~
\log\psi)$, Eq.\,(\ref{eq:lpsi}) represents
a curve where two asymptotes exist:
\begin{leftsubeqnarray}
\slabel{eq:pas0}
&& \log\psi=\log\Delta_{vir}-\log\nu_{mas}+
3\log\xi_{vir} \nonumber \\
&& -\gamma\log\xi_{vir}-\gamma\log\eta~~;\qquad
\xi_{vir}\eta\ll1~~; \\
\slabel{eq:pasi}
&& \log\psi=\log\Delta_{vir}-\log\nu_{mas}+
3\log\xi_{vir} \nonumber \\
&& -\beta\log\xi_{vir}-\beta\log\eta~~;\qquad
\xi_{vir}\eta\gg1~~;
\label{seq:pas}
\end{leftsubeqnarray}
which meet at the point $(\log\eta,\log\psi)
=[\log(r_0/r_{vir}),\log(\rho_0/\rho_h)]$.
For further details, see CMV.

\subsection{Analytical and geometrical properties
of GPL density profiles}\label{geGPL}

The change of coordinates:
\begin{equation}
\label{eq:xy}
x=\log\eta~~;\qquad y=\log\psi~~;
\end{equation}
makes the expressions of the curve and the
asymptotes, defined by Eqs.\,(\ref{eq:lpsi}),
(\ref{eq:pas0}), and (\ref{eq:pasi}), 
translate into: 
\begin{lefteqnarray}
\label{eq:psc}
&& y=y_C-\gamma(x-x_C)-\chi\log\{1+\exp_{10}
[\alpha(x-x_C)]\}~~; \\
\label{eq:p0c}
&& y=y_C-\gamma(x-x_C)~~;\qquad x\ll x_C~~; \\
\label{eq:pic}
&& y=y_C-\beta(x-x_C)~~;\qquad x\gg x_C~~;
\end{lefteqnarray}
where, in general, $\exp_ux=u^x$, and $\exp
x={\rm e}^x$, according to the standard
notation.   The point,
${\sf C}(x_C,y_C)$, represents the
intersection of the asymptotes in the
plane $({\sf O}xy)$, and the expressions
of the coordinates are (CMV):
\begin{lefteqnarray}
\label{eq:xC}
&& x_C=-\log\xi_{vir}=-\log\frac{r_{vir}}{r_0}=
\log\frac{r_0}{r_{vir}}~~; \\
\label{eq:yC}
&& y_C=\log\frac{\Delta_{vir}\xi_{vir}^3}{\nu_
{mas}}=\log\frac{\rho_0}{\rho_h}~~;
\end{lefteqnarray}
which allow the following expression of the
scaled mass:
\begin{equation}
\label{eq:anxy}
\log\nu_{mas}=\log\Delta_{vir}-3x_C-y_C~~.
\end{equation}
The vertical intercept of the curve, $b$,
and the asymptotes, $b_\gamma$ and $b_
\beta$, may be determined through the
combination of Eqs.\,(\ref{eq:psc}), (\ref
{eq:p0c}), (\ref{eq:pic}), and (\ref
{eq:anxy}).   The result is:
\begin{lefteqnarray}
\label{eq:bc}
&& b=y_C+\gamma x_C-\chi\log\left[1+\exp_{10}(-\alpha
x_C)\right]~~; \\
\label{eq:bgac}
&& b_\gamma=y_C+\gamma x_C~~; \\
\label{eq:bbac}
&& b_\beta=y_C+\beta x_C~~;
\end{lefteqnarray}
for further details, see CMV.
                             
It can be proved (CMV) that for arbitrarily
chosen SPH density profile, cosmological
model, and GPL density profile, there is
a one-to-one correspondence between analytical
parameters, $(r_0,\rho_0,\alpha,\beta,\gamma)$,
and geometrical parameters, $(x_C,y_C,b,b_\beta,
b_\gamma)$, in the sense that either set
univocally determines a GPL density profile.
Owing to Eqs.\,(\ref{eq:xC}), and (\ref
{eq:yC}), the set $(x_C,y_C,\alpha,\beta,
\gamma)$ may also be used to this respect.

\section{An analytical RFSM5 method}
\label{RFSM5}

Let $(x_i,y_i)$ be coordinates of a generic
point of a logarithmic SDH density profile,
and $[x_i,y(x_i)]$ be their counterparts
related to a fitting, GPL density profile.
Owing to Eq.\,(\ref{eq:psc}), the corresponding,
logarithmic absolute residual, is:
\begin{equation}
\label{eq:Ri}
R_i=(y_i-y_C)+\gamma(x_i-x_C)+\frac{\beta-
\gamma}\alpha\log\{1+\exp_{10}[\alpha(x_i-
x_C)]\}~~;
\end{equation}
where the explicit expression of the exponent,
$\chi$, has been used, according to Eq.\,(\ref
{eq:GPL}).

With regard to a SDH density profile, a RFSM5
(Residual Function Sum Minimization within a
5-dimension hyperspace) method consists in
the minimization of the sum - over the range
defined by Eq.\,(\ref{eq:doeta}) - of an
assigned residual function (e.g., square,
$R^2$, or absolute value, $\vert R\vert$),
within a selected domain of a 5-dimension
hyperspace, $({\sf O}~x_C~y_C~\alpha~\beta
~\gamma)$,
or any equivalent choice of parameters.   A
numerical procedure has been used in CMV,
where the sum of absolute values and squares
of residuals was performed on $10^5$ nodes
of a hypergrid within the above mentioned
5-dimension hyperspace.   On the contrary,
an analytical procedure shall be exploited
here.

In general, let $v=F(u_1,u_2,...,u_N)$ be
a function which is defined in a domain of
a $N$-dimension hyperspace, $({\sf O}u_1u_2
...u_N)$, and continuous therein.   Then
a necessary and sufficient condition for a
point, ${\sf P}^\ast(u_1^\ast,u_2^\ast,...,
u_N^\ast)$, belonging to the domain, to
be an extremal mimimum, is that the
coordinates of the point under discussion
make (i) a solution of the system, where
all the first partial derivatives equal
zero:
\begin{equation}
\label{eq:sid0}
\left(\frac{\partial F}{\partial u_j}\right)
_{{\sf P}^\ast}=0~~;
\qquad1\le j\le N~~;
\end{equation}
and (ii) the Hessian determinant positive:
\begin{equation}
\label{eq:Hesp}
\left\vert\left(\frac{\partial^2F}{\partial
u_j\partial u_k}\right)_{{\sf P}^\ast}\right
\vert>0~~;\qquad1\le j\le N~~;\qquad1\le k
\le N~~;
\end{equation}
in the more general case where at least one
first partial derivative, $(\partial F/
\partial u_j)_{{\sf P}^\ast}$, does not
exist, Eq.\,(\ref{eq:sid0}) makes a necessary
condition for a point, ${\sf P}^\ast(u_1^\ast,
u_2^\ast,...,$ $u_N^\ast)$, to be an extremum
point of minimum.   For further details see e.g.,
Smirnov (1969, Chap.\,V, \S\S\,22-26).

Owing to the discontinuity of the first
derivative of the absolute value, for null
arguments, which makes a solution of the
problem much more difficult, our attention
shall be limited to the sum of square
residuals in the 5-dimension hyperspace
of interest:
\begin{equation}
\label{eq:sRiq}
F(x_C,y_C,\alpha,\beta,\gamma)=\sum_{i=1}^nR_i^2~~;
\end{equation}
where the $i$-th residual, $R_i$, is
defined by Eq.\,(\ref{eq:Ri}).
Accordingly, Eq.\,(\ref{eq:sid0})
reduces to:
\begin{equation}
\label{eq:sd50}
\left(\frac{\partial F}{\partial u_j}\right)_
{{\sf P}^\ast}=2\sum_{i=1}^n\left(R_i\frac
{\partial R_i}{\partial u_j}\right)_{{\sf P}^
\ast}=0~~;\qquad1\le j\le5~~;
\end{equation}
where $u_1=x_C$, $u_2=y_C$, $u_3=\alpha$,
$u_4=\beta$, $u_5=\gamma$.

The combination of Eqs.\,(\ref{eq:Ri}),
(\ref{eq:sRiq}), and (\ref{eq:sd50}), where the first
derivatives are written explicitly,
after long but stimulating algebra
yields three alternative expressions
of the exponent, $\gamma$, as:
\begin{leftsubeqnarray}
\slabel{eq:gayt}
&& \gamma=\gamma_1(x_C,\alpha)=-\frac
{\upsilon_s(x,s)\upsilon_s(y,t)-
\upsilon_s(s,t)\upsilon_s(x,y)}
{\upsilon_s(x,s)\upsilon_s(x,t)-
\upsilon_s(s,t)\upsilon_s(x,x)}~~; \\
\slabel{eq:gays}
&& \gamma=\gamma_2(x_C,\alpha)=-\frac
{\upsilon_s(x,s)\upsilon_s(y,s)-
\upsilon_s(s,s)\upsilon_s(x,y)}
{\upsilon_s(x,s)\upsilon_s(x,s)-
\upsilon_s(s,s)\upsilon_s(x,x)}~~; \\
\slabel{eq:gayxt}
&& \gamma=\gamma_3(x_C,\alpha)=-\frac
{\upsilon_s(x,s)\upsilon_s(y,xt)-
\upsilon_s(s,xt)\upsilon_s(x,y)}
{\upsilon_s(x,s)\upsilon_s(x,xt)-
\upsilon_s(s,xt)\upsilon_s(x,x)}~~;
\label{seq:gay}
\end{leftsubeqnarray}
where, in general:
\begin{leftsubeqnarray}
\slabel{eq:bsa}
&& \upsilon_s(a,b)=\overline{ab}-
\overline{a}\overline{b}~~; \\
\slabel{eq:bsb}
&& \overline{c}=\frac1n\sum_{i=1}
^nc_i~~;\qquad c=a,b,ab~~;
\label{seq:bs}
\end{leftsubeqnarray}
and the remaining variables, $s_i$
and $t_i$, are defined as:
\begin{leftsubeqnarray}
\slabel{eq:si}
&& s_i=\log(1+w_i)~~; \\
\slabel{eq:ti}
&& t_i=\frac{w_i}{1+w_i}~~; \\
\slabel{eq:wi}
&& w_i=\exp_{10}[\alpha(x_i-x_C)]~~;
\label{seq:stw}
\end{leftsubeqnarray}
where, in addition:
\begin{leftsubeqnarray}
\slabel{eq:dwa}
&& \frac{\partial w_i}{\partial\alpha}=
w_i(x_i-x_C)\ln10~~; \\
\slabel{eq:dwx}
&& \frac{\partial w_i}{\partial x_C}=
-w_i\alpha\ln10~~;
\label{seq:dw}
\end{leftsubeqnarray}
for a formal demonstration, see Appendix A.

In a 3-dimension space, $({\sf O}~x_C~\alpha~
\gamma)$, Eqs.\,(\ref{seq:gay}) represent
three surfaces.   Let $(x_C^\ast~\alpha^\ast~
\gamma^\ast)$ be an intersection point of
the above mentioned surfaces.   The
remaining parameters, $\beta^\ast$ and
$y_C^\ast$, may be calculated using
Eqs.\,(\ref{eq:Rx20}) and (\ref{eq:eR0}),
respectively.   The resulting point in
a 5-dimension space, $({\sf O}~x_C~y_C~
\alpha~\beta~\gamma)$, let it be ${\sf P}^
\ast\equiv
(x_C^\ast,y_C^\ast,\alpha^\ast,\beta^\ast,
\gamma^\ast)$, satisfies the necessary
condition for the occurrence of an extremal
minimum, expressed by Eqs.\,(\ref{seq:msR}).
Then a key role is played by the intersections
of the above mentioned, three surfaces.

To this regard, the following algorithm is
used.   First two pairs of surfaces are
considered, let they be $(\gamma_1,\gamma_2)$
and $(\gamma_1,\gamma_3)$, then the
intersecting lines related to each pair
are simultaneously determined.   If the
intersecting lines exhibit an intersection
point, then the necessary condition for the
occurrence of an extremal minimum is
satisfied at that point.   In the meantime,
it can be numerically tested if the
intersection point is, in fact, a minimum,
thus avoiding the calculation of the
Hessian determinant.   On the contrary,
nothing can be said about the existence
of other extremal minima, and the
coincidence of the point under consideration
with an absolute minimum.

Finding no intersection point does not
necessarily mean absence of extremal
minima: it could imply that the algorithm
used here is insufficient to this respect.
If it is the case, a non extremal minimum
is selected in the subdomain where the
algorithm works.

\section{An application to SDHs on the scale of
cluster of galaxies}\label{ara}

Using an analytical RFSM5 method, GPL
density profiles are fitted to a sample
of 17 SDH density profiles, on the scale
of cluster of galaxies within a flat
$\Lambda$CDM cosmology (Rasia et al.
2004), as already done in CMV.
The values of the cosmological
parameters used therein are: $\Omega_\Lambda
=0.7$; $\Omega_m=0.3$; $\Omega_b=0.03$;
$h=0.7$; $\sigma_8=0.9$; where the symbols
have their usual captions (e.g., Klypin et
al. 2001; Bullock et al. 2001) and, in
particular, the indices $m$ and $b$ denote
all matter (dark + baryons) and baryons only,
respectively.   For a detailed discussion on
the computer code, initial conditions,
resolution issues, and the way of finding the
halo centre, see Rasia et al. (2004) and
further references therein.

Simulations include both dark and baryonic
matter, but only the former is relevant to 
the aim of the current paper.   Accordingly,
the baryonic matter shall not be considered,
and all the parameters shall be intended as
related to the dark matter halo.

The definition of the virialized region
within each halo, via Eq.\,(\ref{eq:Mvir}),
needs the knowledge of the critical
overdensity at virialization,
$\Delta_{vir}$.   With regard to total
(dark + baryonic) matter, it depends on the
cosmological model (e.g., Bullock et al. 2001)
and, in the case under discussion, $(\Delta_
{vir})_m=323$ at $z=0$, where all the sample
haloes may safely be thought of as virialized
(Rasia et al. 2004).

If only the dark matter is considered, then
$(\Delta_{vir})_{cdm}=\Delta_{vir}=\zeta(\Delta_
{vir})_m$, where $\zeta$ is the fraction of dark
matter in each density perturbation, and
averaging over the whole sample yields $\zeta=
0.907$ (Rasia et al. 2004).   Accordingly, the value:
\begin{equation}
\label{eq:Dvir}
\Delta_{vir}=0.907\cdot323=292.961~~;
\end{equation}
can be used to an acceptable extent\footnote%
{The above value of the critical overdensity
was deduced from an earlier, unpublished
version of Rasia et al. (2004).   It is
slightly different from $\Delta_{vir}=
0.903\times323.7625=292.3576$ deduced from
the current version, which appeared when
all the calculations of this paper
were performed.   As
the relative difference amounts to about
0.2\%, the calculations were not repeated
using the latter value of the critical
overdensity.}.%

Owing to the random criterion used for the
selection (Rasia et al. 2004), sample haloes
are characterized by varying dynamical
properties: at the present time, some are
more relaxed, while others are dynamically
perturbed.   The surrounding environment can
also be quite different: some clusters are
more isolated, while others are interacting
with the surrounding cosmic web.   Accordingly,
the related results shall be representative of
an averaged cluster, in an averaged environment
and dynamical configuration.

\subsection{SDH density profiles}
\label{SSDH}

The main features of sample haloes at
$z=0$, are listed in Tab.\,\ref{t:SDH}.
\begin{table}
\begin{tabular}{cccrrrr}
\hline
\hline
\multicolumn{1}{c}{case} &
\multicolumn{1}{c}{run} &
\multicolumn{1}{c}{type}  &
\multicolumn{1}{r}{$\phantom{28}N\phantom{74}$}  &
\multicolumn{1}{r}{$r_{vir}
$} &
\multicolumn{1}{r}{$M_{vir}
$} &
\multicolumn{1}{r}{$M_{vir}^\prime
$} \\
\hline
\phantom{1}1 & $S01.02$ & $R$ & 282574 & 1953 &  76330 &  76040 \\
\phantom{1}2 & $S02.10$ & $M$ & 278569 & 2305 & 125400 & 125010 \\
\phantom{1}3 & $S02.11$ & $M$ &  85159 & 1553 &  38340 &  38234 \\
\phantom{1}4 & $S03.05$ & $R$ & 294373 & 2347 & 132500 & 131970 \\
\phantom{1}5 & $S04.01$ & $R$ & 179681 & 1991 &  80820 &  80565 \\
\phantom{1}6 & $S04.07$ & $R$ & 146386 & 1860 &  65850 &  65686 \\
\phantom{1}7 & $S05.02$ & $M$ & 318653 & 2197 & 108700 & 108249 \\
\phantom{1}8 & $S06.01$ & $M$ & 427583 & 2470 & 153800 & 153825 \\
\phantom{1}9 & $S06.03$ & $M$ & 166855 & 1796 &  60020 &  59136 \\
10           & $S07.01$ & $R$ & 275259 & 1691 &  49600 &  49359 \\
11           & $S07.03$ & $R$ & 158345 & 1407 &  28530 &  28433 \\
12           & $S08.01$ & $R$ & 190453 & 1884 &  68600 &  68262 \\
13           & $S08.04$ & $R$ & 101482 & 1529 &  36560 &  36489 \\
14           & $S09.03$ & $R$ & 159330 & 1913 &  71690 &  71463 \\
15           & $S09.14$ & $R$ & 107229 & 1675 &  48250 &  47971 \\
16           & $S10.03$ & $R$ &  58734 & 1524 &  36060 &  36132 \\
17           & $S10.07$ & $R$ &  71937 & 1628 &  44170 &  44045 \\
\hline
\end{tabular}
\caption{Main features of sample haloes at
$z=0$.   Column captions: 1 - case; 2 -
computer run;
3 - type ($R$ - safely relaxed; $M$ - safely
a major merger occurring); 4 - number of dark
matter particles within the virial radius;
5 - virial radius ($h^{-1}~$kpc); 6 - virial
mass ($h^{-1}10^{10}{\rm m}_\odot$); 7 - virial
mass deduced from Eq.\,(\ref{eq:Mvir}).   Both 
virial radii and virial masses are normalized
to the dimensionless Hubble parameter at the
current time, $h$.}
\label{t:SDH}
\end{table}
The mass, $M_{vir}$, has been taken from
Rasia et al. 2004, while the mass, $M_{vir}^\prime$,
has been deduced from Eq.\,(\ref{eq:Mvir}).
The apparent discrepancy 
between $M_{vir}$ and $M_{vir}^\prime$
is owing to two different sources.
First, a systematic contribution takes
origin from the uncertainty on $\Delta
_{vir}$ and, second, a random contribution
arises from the uncertainty on $r_{vir}$.
An additional random contribution is
related to averaging the fraction of
dark matter over the whole sample,
with regard to Eq.\,(\ref{eq:Mvir}).

The relative difference, $\vert1-M_{vir}
^\prime/M_{vir}\vert$, is less then one
percent in all cases except 9, where it
is less than one and half percent.   The
virial mass can be evaluated, to a good
extent, by use of Eq.\,(\ref{eq:Mvir}),
taking the virial radius from the results
of simulations.   It is worth mentioning
that the RFSM5 method is independent of
the value of the virial mass, while a
change in virial radius makes SDH density
profiles systematically shift along the
horizontal axis in the logarithmic plane
$({\sf O}~\log\eta~\log\psi)$.

The mean SDH density profile 
is obtained by averaging over the whole
set the values related to each logarithmic
radial bin, in the range of interest, 
expressed by Eq.\,(\ref{eq:doeta}).     
The value of the critical overdensity
at virialization, $\Delta_{vir}$,
is fixed by Eq.\,(\ref{eq:Mvir}), then a
single free parameter remains: the virial
radius, $r_{vir}$, which allows the
calculation of the virial mass, $M_{vir}$.
To this aim, first the virial radius is
averaged over the whole sample, using
the data listed in Tab.\,\ref{t:SDH},
and then the mean value of the virial 
radius is inserted into Eq.\,(\ref{eq:Mvir}).
The result is:
\begin{equation}
\label{eq:rMvm}
\overline{r}_{vir}=1866~h^{-1}~{\rm kpc}~~;
\qquad\overline{M}_{vir}=66330~h^{-1}~10^{10}
{\rm m}_\odot~~;
\end{equation}
and the application of a RFSM5 method
yields best fitting GPL density profiles,
with radius and mass equal to $\overline{r}_
{vir}$ and $\overline{M}_{vir}$, respectively.

A mean virial radius has been preferred
in place of a mean concentration (Rasia et al. 2004)
for the following reasons.   First,
virial radii are independent of GPL
density profiles, contrary to concentrations,
or velocity profiles, which should be
calculated for any choice of the fitting
profile.   Second, the range in virial
radius, $1407\le r_{vir}/(h^{-1}{\rm kpc})\le2470$,
corresponds to relative errors of about 
25\% and 33\%, respectively, with regard
to a mean value, $\overline{r}_{vir}=1866~h^{-1}$
kpc.   On the other hand, the range in
concentration (calculated for NFW density
profiles), $5\le\xi_{vir}\le10$,
corresponds to relative errors of about
32\% and 37\%, respectively, with regard
to a mean value, $\overline{\xi}_{vir}=7.2976$.
Then the average of the virial radius
should be preferred to this respect.

Having in our hands a sample of SDH
density profiles, listed in
Tab.\,\ref{t:SDH}, together with the
averaged virial radius, and related
virial mass, expressed by
Eqs.\,(\ref{eq:rMvm}), we are left with
the search of a best fitting GPL density
profile, which minimizes the sum of
square residuals.  To this aim, the
analytical RFSM5 method formulated in
section \ref{RFSM5}, shall be used.

\subsection{Results}
\label{resu}

As outlined in section \ref{RFSM5}, an
analytical RFSM5 method has been applied
to each sample halo, listed in Tab.\,\ref
{t:SDH}, and to the related, mean SDH
density profile, which has been defined
above.   The values of
the exponents, $\alpha$, $\chi$, $\beta$,
$\gamma$, the scaled radius, $\xi_{vir}$,
the scaling radius, $r_0$, the scaling
density, $\rho_0$, and the sum of absolute 
value residuals, $\sum\vert R_i\vert$, and 
square residuals, $\sum R_i^2$, at the 
extremal or fiducial minimum,  are listed
in Tabs.\,\ref{t:numa} and \ref{t:numb},
respectively,
for both analytical (current paper,
upper lines) and numerical (CMV, lower
lines) RFSM5 methods.
The following conclusions are deduced.
\begin{table}
\begin{tabular}{ccllllc}
\hline
\hline
\multicolumn{1}{c}{case} &
\multicolumn{1}{c}{type} &
\multicolumn{1}{c}{$\alpha$} &
\multicolumn{1}{c}{$\chi$} & 
\multicolumn{1}{c}{$\beta$} &
\multicolumn{1}{c}{$\gamma$} &
\multicolumn{1}{c}{class} \\
\hline
\phantom{1}1 & I & \phantom{1}0.76000 & 21.296 & 17.533 & 1.3480
  & IMP \\
  & & \phantom{1}0.85979 & \phantom{1}2.7291 & \phantom{1}3.4893
  & 1.1429 & \\
\phantom{1}2 & M & \phantom{1}8.96000 & \phantom{1}0.081928
  & \phantom{1}2.1484 & 1.4143 & BMP \\
  & & \phantom{1}0.70267 & \phantom{1}2.6771 & \phantom{1}2.8238
  & 0.94262 & \\
\phantom{1}3 & M & \phantom{1}7.60000 & \phantom{1}0.072676
 & \phantom{1}2.1063 & 1.5540 & IMP \\
 & & \phantom{1}0.50594 & \phantom{1}3.6824 & \phantom{1}3.0631
 & 1.2000 & \\
\phantom{1}4 & I & \phantom{1}3.9760 & \phantom{1}0.14895
 & \phantom{1}1.9593 & 1.3671 & IMP \\
 & & \phantom{1}0.40157 & \phantom{1}4.8960 & \phantom{1}2.9564
 & 0.99023 & \\
\phantom{1}5 & I & \phantom{1}5.4644 & \phantom{1}0.12476
 & \phantom{1}2.3843 & 1.7026 & EMP \\
 & & \phantom{1}0.59295 & \phantom{1}3.2455 & \phantom{1}3.1244
 & 1.2000 & \\
\phantom{1}6 & I & 19.787 & \phantom{1}0.027126 & \phantom{1}2.0238
 & 1.4870 & EMP \\
 & & \phantom{1}0.50437 & \phantom{1}2.8516 & \phantom{1}2.5152
 & 1.0769 & \\
\phantom{1}7 & M & \phantom{1}7.5000 & \phantom{1}0.13191
 & \phantom{1}2.2709 & 1.2816 & IMP \\
 & & \phantom{1}0.60451 & \phantom{1}4.1915 & \phantom{1}3.5715
 & 1.0377 & \\
\phantom{1}8 & M & \phantom{1}0.42000 & 12.214 & \phantom{1}5.7327
 & 0.60284 & IMP \\
 & & \phantom{1}0.66861 & \phantom{1}3.9138 & \phantom{1}3.4748 
 & 0.85806 & \\
\phantom{1}9 & M & \phantom{1}0.77830 & \phantom{1}2.1159
 & \phantom{1}2.7751 & 1.1283 & EMP \\
 & & \phantom{1}0.58477 & \phantom{1}3.7717 & \phantom{1}3.2083
 & 1.0027 & \\
10 & I & 13.5000 & \phantom{1}0.045375 & \phantom{1}1.9470
 & 1.3344 & IMP \\
 & & \phantom{1}0.43019 & \phantom{1}3.7611 & \phantom{1}2.7372
 & 1.1192 & \\
11 & I & \phantom{1}3.3124 & \phantom{1}0.28130 & \phantom{1}2.5344
 & 1.6026 & EMP \\
 & & \phantom{1}0.69931 & \phantom{1}3.3055 & \phantom{1}3.4625
 & 1.1509 & \\
12 & I & \phantom{1}2.9758 & \phantom{1}0.35736 & \phantom{1}2.0978
 & 1.0344 & EMP \\
 & & \phantom{1}0.67451 & \phantom{1}3.0786 & \phantom{1}2.7795
 & 0.70312 & \\
13 & I & \phantom{1}2.1590 & \phantom{1}0.53250 & \phantom{1}2.5151
 & 1.3655 & EMP \\
 & & \phantom{1}0.68562 & \phantom{1}3.3840 &\phantom{1}3.1595
 & 0.83933 & \\
14 & I & \phantom{1}0.57290 & \phantom{1}6.0145 & \phantom{1}4.3163
 & 0.87060 & EMP \\
 & & \phantom{1}0.81937 & \phantom{1}2.4907 & \phantom{1}2.9870
 & 0.94621 & \\
15 & I & \phantom{1}3.8132 & \phantom{1}0.22865 & \phantom{1}2.2556
 & 1.3837 & EMP \\
 & & \phantom{1}0.70472 & \phantom{1}2.8497 & \phantom{1}2.8694
 & 0.86117 & \\
16 & I & \phantom{1}1.4000 & \phantom{1}1.2555 & \phantom{1}3.0649
 & 1.3072 & IMP \\
 & & \phantom{1}0.74414 & \phantom{1}3.9381 & \phantom{1}3.8674
 & 0.93684 & \\
17 & I & 11.0000 & \phantom{1}0.039069 & \phantom{1}1.9504
 & 1.5207 & IMP \\
 & & \phantom{1}0.44450 & \phantom{1}4.0781 & \phantom{1}2.8351
 & 1.0224 & \\
ADP & & \phantom{1}0.29000 & 14.846 & \phantom{1}4.4590 & 1.5360
 & IMP \\
 & & \phantom{1}0.56832 & \phantom{1}4.0722 & \phantom{1}3.3143
 & 1.0000 & \\
\hline\hline
\end{tabular}
\caption{Exponents of GPL density profiles which
minimize the sum of square residuals, using (i)
an analytical RFSM5 method (current paper, upper
lines), and (ii) a numerical RFSM5 method (CMV,
lower lines), with regard to 17 sample haloes
listed in Tab.\,\ref{t:SDH}, together with the
averaged density profile (ADP).   Type captions:
I - safely relaxed; M - safely a major merger
occurring.   Class captions: EMP - extremal
minimum point; IMP - minimum point occurring
within the subdomain interested by the
application of the method, where a necessary
condition for the existence of an extremal
point is not satisfied; BMP - minimum point
occurring on the boundary of the subdomain
interested by the application of the method,
where a necessary condition for the existence
of an extremal point is not satisfied.}
\label{t:numa}
\end{table}
\begin{table}
\begin{tabular}{cclllllc}
\hline
\hline
\multicolumn{1}{c}{case} &
\multicolumn{1}{c}{type} &
\multicolumn{1}{c}{$\xi_{vir}$} &
\multicolumn{1}{c}{$r_0$} &
\multicolumn{1}{c}{$\rho_0$} &
\multicolumn{1}{c}{$\sum\vert R_i\vert$} &
\multicolumn{1}{c}{$\sum R_i^2$} &
\multicolumn{1}{c}{class} \\
\hline
\phantom{1}1 & I & 8.5258~E$-$2 & 3.2724~E+4 &
1.4808~E$-$7 & 2.7556~E$-$1 & 5.3397~E$-$3 & IMP \\
 & & 3.6308~E+0 & 7.6843~E+2 & 4.5733~E$-$5 &
4.3189~E$-$1 & 1.2843~E$-$2 & \\
\phantom{1}2 & M & 1.0235~E+1 & 3.2172~E+2 &
5.1755~E$-$5 & 6.4689~E$-$1 & 3.5769~E$-$2 & BMP \\
 & & 7.1450~E+0 & 4.6086~E+2 & 1.4462~E$-$4 &
6.7590~E$-$1 & 5.0984~E$-$2 & \\
\phantom{1}3 & M & 2.0099~E+1 & 1.1038~E+3 &
2.0053~E$-$4 & 6.9707~E$-$1 & 7.5840~E$-$2 & IMP \\
 & & 4.4668~E+0 & 4.9668~E+2 & 1.1755~E$-$4 &
8.2813~E$-$1 & 5.9798~E$-$2 & \\
\phantom{1}4 & I & 2.3393~E+1 & 1.4333~E+2 &
2.0029~E$-$4 & 7.3567~E$-$1 & 5.1734~E$-$2 & IMP \\
 & & 4.1115~E+0 & 8.1548~E+2 & 2.1391~E$-$4 &
7.7711~E$-$1 & 4.4834~E$-$2 & \\
\phantom{1}5 & I & 7.0536~E+0 & 4.0324~E+2 &
3.0094~E$-$5 & 3.9790~E$-$1 & 1.3080~E$-$2 & EMP \\
 & & 4.7863~E+0 & 5.9426~E+2 & 1.0721~E$-$4 &
4.5994~E$-$1 & 1.5935~E$-$2 & \\
\phantom{1}6 & I & 2.0441~E+1 & 1.2999~E+2 &
1.7204~E$-$4 & 3.7559~E$-$1 & 1.2647~E$-$2 & EMP \\
 & & 1.6501~E+1 & 1.6103~E+2 & 7.3632~E$-$4 &
6.2224~E$-$1 & 3.3055~E$-$2 & \\
\phantom{1}7 & M & 2.1719~E+1 & 1.4451~E+2 &
3.3786~E$-$4 & 7.9368~E$-$1 & 4.8776~E$-$2 & IMP \\
 & & 4.2267~E+0 & 7.4256~E+2 & 1.7070~E$-$4 &
6.1250~E$-$1 & 2.9137~E$-$2 & \\
\phantom{1}8 & M & 5.4741~E$-$1 & 6.4459~E+3 &
1.9954~E$-$4 & 2.4863~E$-$1 & 4.5390~E$-$3 & IMP \\
 & & 4.1687~E+0 & 8.4645~E+2 & 1.3190~E$-$4 &
2.2446~E$-$1 & 4.9367~E$-$3 & \\
\phantom{1}9 & M & 9.4211~E+0 & 2.7234~E+2 &
2.0100~E$-$4 & 2.6901~E$-$1 & 5.7462~E$-$3 & EMP \\
 & & 5.6234~E+0 & 4.5626~E+2 & 2.2399~E$-$4 &
2.8092~E$-$1 & 6.1089~E$-$3 & \\
10 & I & 3.3231~E+1 & 7.2696~E+2 & 3.8804~E$-$4 &
4.6024~E$-$1 & 2.4169~E$-$2 & IMP \\
 & & 6.6374~E+0 & 3.6395~E+2 & 2.5132~E$-$4 &
5.0765~E$-$1 & 2.0731~E$-$2 & \\
11 & I & 6.9351~E+0 & 2.8983~E+2 & 3.4633~E$-$5 &
3.3590~E$-$1 & 9.4062~E$-$3 & EMP \\
 & & 3.7154~E+0 & 5.4100~E+2 & 6.7645~E$-$5 &
4.5841~E$-$1 & 1.7674~E$-$2 & \\
12 & I & 1.6379~E+1 & 1.6432~E+2 & 1.2941~E$-$4 &
4.8093~E$-$1 & 2.8344~E$-$2 & EMP \\
 & & 9.1201~E+0 & 3.6319~E+2 & 2.8856~E$-$4 &
8.1443~E$-$1 & 5.0487~E$-$2 & \\
13 & I & 8.0514~E+0 & 2.4964~E+2 & 5.2008~E$-$5 &
6.1515~E$-$1 & 3.2976~E$-$2 & EMP \\
 & & 6.8234~E+0 & 3.2012~E+2 & 2.5718~E$-$4 &
5.9097~E$-$1 & 3.6570~E$-$2 & \\
14 & I & 1.3437~E+0 & 2.0338~E+3 & 3.9969~E$-$5 &
3.8299~E$-$1 & 1.1169~E$-$2 & EMP \\
 & & 5.3088~E+0 & 5.1477~E+2 & 7.7666~E$-$5 &
4.1735~E$-$1 & 1.2273~E$-$2 & \\
15 & I & 8.9712~E+0 & 2.6673~E+2 & 4.6219~E$-$5 &
4.9331~E$-$1 & 2.0331~E$-$2 & EMP \\
 & & 7.6913~E+0 & 3.1111~E+2 & 1.9963~E$-$4 &
6.0948~E$-$1 & 2.8821~E$-$2 & \\
16 & I & 5.2999~E+0 & 4.1079~E+2 & 3.8564~E$-$5 &
6.4586~E$-$1 & 2.9673~E$-$2 & IMP \\
 & & 3.7154~E+0 & 5.8599~E+2 & 1.1226~E$-$4 &
4.4819~E$-$1 & 1.5259~E$-$2 & \\
17 & I & 1.4047~E+1 & 1.6556~E+2 & 7.2820~E$-$5 &
7.4326~E$-$1 & 7.3645~E$-$2 & IMP \\
 & & 4.6774~E+0 & 4.9723~E+2 & 1.5144~E$-$4 &
7.1010~E$-$1 & 5.7718~E$-$2 & \\
ADP & & 2.2073~E+0 & 1.2077~E+3 & 5.9044~E$-$6 &
1.1027~E$-$1 & 1.1783~E$-$3 & IMP \\
 & & 3.6308~E+0 & 7.3422~E+2 & 1.0238~E$-$4 & 
1.3262~E$-$1 & 1.4983~E$-$3 \\
\hline\hline
\end{tabular}
\caption{Scale parameters of GPL density profiles which
minimize the sum of square residuals, using (i)
an analytical RFSM5 method (current paper, upper
lines), and (ii) a numerical RFSM5 method (CMV,
lower lines), with regard to 17 sample haloes
listed in Tab.\,\ref{t:SDH}, together with the
averaged density profile (ADP).   The scaling
radius, $r_0$,  and scaling density, $\rho_0$,
are expressed in kpc and $10^{10}{\rm m}_\odot
/{\rm kpc}^3$, respectively.   Other captions as
in Tab.\,\ref{t:numa}}.
\label{t:numb}
\end{table}
\begin{description}
\item[\rm{(i)}\hspace{3.5mm}] A necessary
condition for the occurrence of an extremal
point, expressed by Eqs.\,(\ref{seq:gay}),
is satisfied for 8 SDHs and is not for the
remaining 9 together with ADP.   In the
former alternative, an extremal minimum
point may safely exist.   In the latter
alternative, the existence of an extremal
minimum point cannot be established, but
only a non extremal minimum may be determined.
This last is related to the GPL density
profile where the distance between intersecting
lines of two distinct pairs of surfaces,
defined by Eqs.\,(\ref{seq:gay}), attains the
lowest value.   The above mentioned, non
extremal minimum, may be located both within
(IMP) and on the boundary (BMP) of the
subdomain where the RFSM5 method, used in
the current paper, works.
\item[\rm{(ii)}~~] The sum of square residuals
is systematically lower using analytical
(current paper) instead of numerical (CMV)
RFSM5 methods, provided an extremal minimum
point occurs.   Accordingly, EMPs may safely
be thought of as absolute minima.   With
regard to non extremal minimum points, the
above result maintains in 4 cases (including
ADP) while the contrary holds for the
remaining 5 cases.   The trend is not
necessarily the same for the sum of absolute
value residuals, as expected.
\item[\rm{(iii)}~] The exponents, $\alpha$,
$\beta$, $\gamma$, range within $0.4<\alpha
<19.8$ ($\alpha=0.2900$ for ADP); $1.9<\beta
<17.6$; $0.6<\gamma<1.8$; and $0.4<\alpha<0.9
$; $2.5<\beta<3.9$; $0.7<\gamma<1.3$; using
analytical and numerical RFSM5 methods,
respectively.   With regard to SDH density
profiles where an EMP occurs, the above
ranges in the former alternative reduce to
$0.5<\alpha<19.8$; $2.0<\beta<4.4$ ($\beta=
4.4590$ for ADP); $0.8<\gamma<1.8$; which
is, only for $\beta$, substantially closer
to their counterparts related to the latter
alternative.   Accordingly, analytical RFSM5
methods yield GPL density profiles where the
range in the exponents is larger than in
their counterparts related to numerical RSFM5
methods.   It provides additional support to
the occurrence of a certain degree of
degeneracy (e.g., Klypin et al. 2001; CMV).
\item[\rm{(iv)}~] The scaled radius, $\xi_
{vir}$, the scaling radius, $r_0$, and the
scaling density, $\rho_0$, range within
$0.08<\xi_{vir}<33.3$; $72<r_0/{\rm kpc}<
32724$; $0.01<10^4\rho_0/(10^{10}{\rm m}_\odot/
{\rm kpc}^3)<3.9$; and $3.6<\xi_{vir}<16.5$;
$161<r_0/{\rm kpc}<847$; $0.4<10^4\rho_0/
(10^{10}{\rm m}_\odot/{\rm kpc}^3)<7.4$; using
analytical and numerical RFSM5 methods,
respectively.   With regard to SDH density
profiles where an EMP occurs, the above
ranges in the former alternative reduce to
$1.3<\xi_{vir}<20.5$; $129<r_0/{\rm kpc}<
2034$; $0.3<10^4\rho_0/(10^{10}{\rm m}_\odot/
{\rm kpc}^3)<2.1$; which is much closer
to their counterparts related to the latter
alternative.
\end{description}

The occurrence of high values of the
exponents, $\alpha$ and $\beta$, related
to (a) lack of an EMP, and (b) lower value
of the sum of square residuals using
analytical instead of numerical RFSM5
methods, as in cases 1 and 10 of Tab.\,\ref
{t:numa}, suggests further investigation
on the best fitting, limiting GPL density
profiles, analysed in subsection \ref{GPL}.

According to Eq.\,(\ref{eq:liai}), GPL
density profiles reduce to a double
power-law (DPL) with exponents equal to
$-\gamma$ and $-\beta$, in the limit 
$\alpha\to\infty$, $\chi\to0$, i.e.
$\gamma$ and $\beta$ finite.   Then
we are left with two exponents instead
of three, and a RFSM4 method has to be
used, within a 4-dimension hyperspace
$({\sf O}~x_C~y_C~\beta~\gamma)$, in
dealing with each sample halo, listed
in Tab.\,\ref{t:SDH}, together with
the related ADP, which has been defined
above.   The values of the exponents,
$\beta$ and $\gamma$, the scaled radius,
$\xi_{vir}$, the scaling radius, $r_0$,
the scaling density, $\rho_0$, the sum
of absolute value residuals, $\sum\vert
R_i\vert$, and the sum of square residuals,
$\sum R_i^2$, are listed in Tab.\,\ref
{t:DPL}.   By comparison with Tabs.\,\ref
{t:numa} and \ref{t:numb}, the following
conclusions are deduced.
\begin{table}
\begin{tabular}{
lllllll
}
\hline
\hline
\multicolumn{1}{c}{$\beta$} &
\multicolumn{1}{c}{$\gamma$} &
\multicolumn{1}{c}{$\xi_{vir}$} &
\multicolumn{1}{c}{$r_0$} &
\multicolumn{1}{c}{$\rho_0$} &
\multicolumn{1}{c}{$\sum\vert R_i\vert$} &
\multicolumn{1}{c}{$\sum R_i^2$}
\\
\hline
2.8057 & 1.6263 & 3.9799~E+0 &
7.1465~E+2 & 1.0852~E$-$5 & 9.9122~E$-$1 & 1.4472~E$-$1 \\
2.1545 & 1.4338 & 9.6963~E+0 &
2.9334~E+2 & 4.8193~E$-$5 & 6.4395~E$-$1 & 3.5625~E$-$2 \\
3.4200 & 1.8638 & 1.9336~E+0 &
1.4710~E+3 & 1.9761~E$-$6 & 9.1288~E$-$1 & 5.1722~E$-$2 \\
3.0400 & 1.7367 & 1.9743~E+0 &
1.4407~E+3 & 2.0941~E$-$6 & 7.0852~E$-$1 & 4.3206~E$-$2 \\
2.3798 & 1.7257 & 6.6950~E+0 &
4.2484~E+2 & 2.7050~E$-$5 & 3.9284~E$-$1 & 1.3590~E$-$2 \\
2.0238 & 1.4900 & 2.0314~E+1 &
1.4001~E+2 & 1.7556~E$-$4 & 3.7440~E$-$1 & 1.2651~E$-$2 \\
2.2659 & 1.3214 & 2.0573~E+1 &
1.3825~E+2 & 2.9438~E$-$4 & 8.5352~E$-$1 & 5.0746~E$-$2 \\
2.3228 & 1.4364 & 8.1896~E+0 &
3.4731~E+2 & 4.1276~E$-$5 & 5.6538~E$-$1 & 2.3083~E$-$2 \\
2.3222 & 1.6425 & 8.0737~E+0 &
3.5229~E+2 & 3.9237~E$-$5 & 4.2369~E$-$1 & 1.4978~E$-$2 \\
1.9452 & 1.3370 & 3.3206~E+1 & 8.5655~E+2 &
3.8561~E$-$4 & 4.6730~E$-$1 & 2.4272~E$-$2 \\
2.4412 & 1.6360 & 7.5301~E+0 & 3.7772~E+2 &
3.7991~E$-$5 & 4.0076~E$-$1 & 1.3061~E$-$2 \\
2.0438 & 1.1446 & 1.5593~E+1 & 1.8241~E+2 &
1.0817~E$-$4 & 6.1143~E$-$1 & 3.3369~E$-$2 \\
2.4638 & 1.5349 & 6.4809~E+0 & 4.3887~E+2 &
2.7799~E$-$5 & 7.0267~E$-$1 & 3.8464~E$-$2 \\
2.2258 & 1.3844 & 8.2336~E+0 & 3.4545~E+2 &
3.7441~E$-$5 & 5.1443~E$-$1 & 2.2151~E$-$2 \\
2.2410 & 1.4449 & 8.1530~E+0 & 3.4886~E+2 &
3.7169~E$-$5 & 5.3382~E$-$1 & 2.2494~E$-$2 \\
2.5215 & 1.5093 & 7.8512~E+0 & 3.6228~E+2 &
4.6051~E$-$5 & 8.7268~E$-$1 & 5.5935~E$-$2 \\
1.9366 & 1.5001 & 1.5659~E+1 & 1.8164~E+2 &
8.8556~E$-$5 & 7.1163~E$-$1 & 7.3474~E$-$2 \\
2.1974 & 1.5207 & 9.7371~E+0 & 2.9211~E+2 &
5.1240~E$-$5 & 4.2918~E$-$1 & 1.3103~E$-$2 \\
\hline\hline
\end{tabular}
\caption{Parameters of double power-law (DPL)
density profiles which minimize the sum of
square residuals, using an analytical RFSM4
method, with regard to 17 sample haloes
listed in Tab.\,\ref{t:SDH} together with the
averaged density profile (ADP).  Other captions
as in Tabs.\,\ref{t:numa}-\ref{t:numb}. Cases,
types, and classes have been omitted to save
space.}
\label{t:DPL}
\end{table}
\begin{description}
\item[\rm{(v)}\hspace{2.5mm}] The sum of
square residuals is systematically larger
with respect to analytical RFSM5 methods,
provided an EMP occurs, as expected.
Concerning non extremal minimum points,
the above result maintains in 6 cases
(including ADP), while the contrary holds
for the remaining 4 cases.   The sum of
square residuals is larger, with respect
to numerical RSFM5 methods, in 11 cases
(including ADP), while the contrary holds
for the remaining 8 cases.   The trend is
not necessarily the same for the sum of
absolute value residuals, as expected.
\item[\rm{(vi)}~] The exponents, $\beta$
and $\gamma$, range within $1.9<\beta<3.5$
and $1.1<\gamma<1.9$, respectively, which
are comparable to their counteparts related
to both analytical (restricted to occurring
EMPs) and numerical RSFM5 methods.
\item[\rm{(vii)}] The scaled radius, $\xi_
{vir}$, the scaling radius, $r_0$, and the
scaling density, $\rho_0$, range within
$1.9<\xi_{vir}<33.3$; $85<r_0/{\rm kpc}<
1471$; and $0.01<10^4\rho_0/(10^{10}{\rm
m}_\odot/{\rm kpc}^3)<3.9$; respectively,
which are comparable to their counterparts
related to both analytical (restricted to
occurring EMPs) and numerical RFSM5 methods.
\end{description}
Then DPL density profiles may be considered
as a rough, but viable approximation in
fitting SDH density profiles.

According to Eqs.\,(\ref{eq:libi}) and
(\ref{eq:lia0}), GPL density profiles
reduce to a simple power-law (SPL) with
slope equal to $-\gamma$, in the limit
$\beta\to+\infty$ where $\alpha$,
$\gamma$, and $(1+\xi^\alpha)^\chi$
remain finite, and $\alpha\to0$ where
$\beta\to\gamma$ and $(1+\xi^\alpha)^
\chi\to2^\chi$.   In any case, we are
left with one exponent instead of three
and one coordinate (the vertical intercept
of SPL density profiles in the logarithmic
plane, $b_\gamma$) instead of two.   Then
a RFSM2 method i.e. a standard least-squares
fit has to be used, within a plane $({\sf O}~
b_\gamma~\gamma)$, in dealing with each sample
halo, listed in Tab.\,\ref{t:SDH}, together
with the related ADP, which has been defined
above.   The values of the exponent, $\gamma$,
the intercept, $b_\gamma$, the sum of absolute
value residuals, $\sum\vert R_i\vert$, and the
sum of square residuals, $\sum R_i^2$, are
listed in Tab.\,\ref{t:SPL}.   By comparison
with Tabs.\,\ref{t:numa} and \ref{t:numb},
the following conclusions are deduced.
\begin{table}
\begin{tabular}{ccllllc}
\hline
\hline
\multicolumn{1}{c}{case} &
\multicolumn{1}{c}{type} &
\multicolumn{1}{c}{$\gamma$} &
\multicolumn{1}{c}{$b_\gamma$} &
\multicolumn{1}{c}{$\sum\vert R_i\vert$} &
\multicolumn{1}{c}{$\sum R_i^2$} &
\multicolumn{1}{c}{class} \\
\hline
\phantom{1}1 & I & 1.9078 & 2.0722 & 2.2846 & 0.35794 & IMP \\
\phantom{1}2 & M & 1.7869 & 2.1141 & 1.8812 & 0.24998 & BMP \\
\phantom{1}3 & M & 1.9499 & 2.0314 & 1.3315 & 0.16573 & IMP \\
\phantom{1}4 & I & 1.8133 & 2.0893 & 1.2889 & 0.12975 & IMP \\
\phantom{1}5 & I & 1.9680 & 2.0322 & 1.5870 & 0.17503 & EMP \\
\phantom{1}6 & I & 1.8764 & 2.0567 & 1.1596 & 0.099672 & EMP \\
\phantom{1}7 & M & 2.0086 & 2.0223 & 2.0516 & 0.32024 & IMP \\
\phantom{1}8 & M & 1.8220 & 2.1132 & 2.1881 & 0.34027 & IMP \\
\phantom{1}9 & M & 1.9351 & 2.0572 & 1.7520 & 0.20064 & EMP \\
10 & I & 1.8576 & 2.0682 & 0.96462 & 0.082826 & IMP \\
11 & I & 1.9644 & 2.0418 & 2.0423 & 0.26923 & EMP \\
12 & I & 1.7228 & 2.1469 & 2.2995 & 0.33084 & EMP \\
13 & I & 1.8695 & 2.0913 & 2.3278 & 0.35789 & EMP \\
14 & I & 1.7519 & 2.1432 & 2.0687 & 0.30842 & EMP \\
15 & I & 1.7901 & 2.1227 & 2.0813 & 0.27796 & EMP \\
16 & I & 1.9358 & 2.0614 & 2.5099 & 0.46469 & IMP \\
17 & I & 1.7813 & 2.0982 & 1.2451 & 0.14339 & IMP \\
ADP & & 1.8531 & 2.0856 & 1.7061 & 0.20209 & IMP \\
\hline\hline
\end{tabular}
\caption{Parameters of single power-law (SPL) density
profiles which minimize the sum of square residuals,
using an analytical RFSM2 method i.e. a standard
least squares fit, with regard to 17 sample haloes
listed in Tab.\,\ref{t:SDH} together with the
averaged density profile (ADP).   Other captions
as in Tabs.\,\ref{t:numa} and \ref{t:numb}.}
\label{t:SPL}
\end{table}
\begin{description}
\item[\rm{(viii)}\hspace{2.5mm}] The sum of
square residuals is systematically larger
with respect to both analytical and numerical
RFSM5 methods.   The same holds also for
the sum of absolute value residuals.
\item[\rm{(ix)}~] The exponent, $\gamma$,
and the intercept, $b_\gamma$, range within
$1.7<\gamma<2.1$ and $2.0<b_\gamma<2.2$, which
is larger and lower, respectively, with regard
to their counteparts related to both analytical
and numerical RSFM5 methods, as expected.
\end{description}
Then SPL density profiles cannot be considered
as an acceptable approximation to GPL density
profiles, in fitting SDH density profiles.

To get further insight, the sum of square residuals
is listed in Tab.\,\ref{t:srq} for all the four
alternatives which have been exploited in the
current section.
\begin{table}
\begin{tabular}{ccllllc}
\hline
\hline
\multicolumn{1}{c}{case} &
\multicolumn{1}{c}{type} &
\multicolumn{1}{c}{GPN} &
\multicolumn{1}{c}{GPA} &
\multicolumn{1}{c}{DPA} &
\multicolumn{1}{c}{SPA} &
\multicolumn{1}{c}{class} \\
\hline
\phantom{1}1 & I & 1.2843~E$-$2 & 5.3397~E$-$3 & 1.4472~E$-$1 &
3.5794~E$-$1 & IMP \\
\phantom{1}2 & M & 5.0984~E$-$2 & 3.5769~E$-$2 & 3.5625~E$-$2 &
2.4998~E$-$1 & BMP \\
\phantom{1}3 & M & 5.9798~E$-$2 & 7.5840~E$-$2 & 5.1722~E$-$2 &
1.6573~E$-$1 & IMP \\
\phantom{1}4 & I & 4.4834~E$-$2 & 5.1734~E$-$2 & 4.3206~E$-$2 &
1.2975~E$-$1 & IMP \\
\phantom{1}5 & I & 1.5935~E$-$2 & 1.3080~E$-$2 & 1.3590~E$-$2 &
1.7503~E$-$1 & EMP \\
\phantom{1}6 & I & 3.3055~E$-$2 & 1.2647~E$-$2 & 1.2651~E$-$2 &
9.9672~E$-$2 & EMP \\
\phantom{1}7 & M & 2.9137~E$-$2 & 4.8776~E$-$2 & 5.0746~E$-$2 &
3.2024~E$-$1 & IMP \\
\phantom{1}8 & M & 4.9367~E$-$3 & 4.5390~E$-$3 & 2.3083~E$-$2 &
3.4027~E$-$1 & IMP \\
\phantom{1}9 & M & 6.1089~E$-$3 & 5.7462~E$-$3 & 1.4978~E$-$2 &
2.0064~E$-$1 & EMP \\
10 & I & 2.0731~E$-$2 & 2.4169~E$-$2 & 2.4272~E$-$2 &
8.2826~E$-$2 & IMP \\
11 & I & 1.7674~E$-$2 & 9.4062~E$-$3 & 1.3061~E$-$2 &
2.6923~E$-$1 & EMP \\
12 & I & 5.0487~E$-$2 & 2.8344~E$-$2 & 3.3369~E$-$2 &
3.3084~E$-$1 & EMP \\
13 & I & 3.6570~E$-$2 & 3.2976~E$-$2 & 3.8464~E$-$2 &
3.5789~E$-$1 & EMP \\
14 & I & 1.2273~E$-$2 & 1.1169~E$-$2 & 2.2151~E$-$2 &
3.0842~E$-$1 & EMP \\
15 & I & 2.8821~E$-$2 & 2.0331~E$-$2 & 2.2494~E$-$2 &
2.7796~E$-$1 & EMP \\
16 & I & 1.5259~E$-$2 & 2.9673~E$-$2 & 5.5935~E$-$2 &
4.6469~E$-$1 & IMP \\
17 & I & 5.7718~E$-$2 & 7.3645~E$-$2 & 7.3474~E$-$2 &
1.4339~E$-$1 & IMP \\
ADP & & 1.4983~E$-$3 & 1.1783~E$-$3 & 1.3103~E$-$2 &
2.0209~E$-$1 & IMP \\
\hline\hline
\end{tabular}
\caption{The minimized sum of square residuals
which provides the best fit to 17 sample haloes
listed in Tab.\,\ref{t:SDH} together with the
averaged density profile (ADP), with regard to
the following alternatives: numerical RFSM5
methods involving GPL density profiles (GPN);
analytical RFSM5 methods involving GPL density
profiles (GPA); analytical RFSM4 methods
involving DPL density profiles (DPA);
analytical RFSM2 methods involving SPL density
profiles i.e. standard least squares method
(SPA).   Other captions
as in Tabs.\,\ref{t:numa} and \ref{t:numb}.}
\label{t:srq}
\end{table}
The values of
the exponents, $\alpha$, $\chi$, $\beta$,
$\gamma$, the scaled radius, $\xi_{vir}$,
the scaling radius, $r_0$, the scaling
density, $\rho_0$, and the sum of absolute 
value residuals, $\sum\vert R_i\vert$, and 
square residuals, $\sum R_i^2$, at the 
extremal or fiducial minimum, are listed
in Tabs.\,\ref{t:qmia} and \ref{t:qmib},
respectively,
with regard to the best fit, among the
ones considered above, for each sample
halo.
\begin{table}
\begin{tabular}{ccllllcc}
\hline
\hline
\multicolumn{1}{c}{case} &
\multicolumn{1}{c}{type} &
\multicolumn{1}{c}{$\alpha$} &
\multicolumn{1}{c}{$\chi$} & 
\multicolumn{1}{c}{$\beta$} &
\multicolumn{1}{c}{$\gamma$} &
\multicolumn{1}{c}{class} &
\multicolumn{1}{c}{method} \\
\hline
\phantom{1}1 & I & \phantom{1}0.76000 & 21.296 & 17.533 & 1.3480
  & IMP & GPA\\
\phantom{1}2 & M & $+\infty$ & \phantom{1}0
  & \phantom{1}2.1545 & 1.4338 & BMP & DPA \\
\phantom{1}3 & M & $+\infty$ & \phantom{1}0
 & \phantom{1}3.4200 & 1.8638 & IMP & DPA \\
\phantom{1}4 & I & $+\infty$ & \phantom{1}0
 & \phantom{1}3.0400 & 1.7367 & IMP & DPA \\
\phantom{1}5 & I & \phantom{1}5.4644 & \phantom{1}0.12476
 & \phantom{1}2.3843 & 1.7026 & EMP & GPA \\
\phantom{1}6 & I & 19.787 & \phantom{1}0.027126 & \phantom{1}2.0238
 & 1.4870 & EMP & GPA \\
\phantom{1}7 & M &
\phantom{1}0.60451 & \phantom{1}4.1915 & \phantom{1}3.5715
 & 1.0377 & IMP & GPN \\
\phantom{1}8 & M & \phantom{1}0.42000 & 12.214 & \phantom{1}5.7327
 & 0.60284 & IMP & GPA \\
\phantom{1}9 & M & \phantom{1}0.77830 & \phantom{1}2.1159
 & \phantom{1}2.7751 & 1.1283 & EMP & GPA \\
10 & I &
\phantom{1}0.43019 & \phantom{1}3.7611 & \phantom{1}2.7372
 & 1.1192 & IMP & GPN \\
11 & I & \phantom{1}3.3124 & \phantom{1}0.28130 & \phantom{1}2.5344
 & 1.6026 & EMP & GPA \\
12 & I & \phantom{1}2.9758 & \phantom{1}0.35736 & \phantom{1}2.0978
 & 1.0344 & EMP & GPA \\
13 & I & \phantom{1}2.1590 & \phantom{1}0.53250 & \phantom{1}2.5151
 & 1.3655 & EMP & GPA \\
14 & I & \phantom{1}0.57290 & \phantom{1}6.0145 & \phantom{1}4.3163
 & 0.87060 & EMP & GPA \\
15 & I & \phantom{1}3.8132 & \phantom{1}0.22865 & \phantom{1}2.2556
 & 1.3837 & EMP & GPA \\
16 & I &
\phantom{1}0.74414 & \phantom{1}3.9381 & \phantom{1}3.8674
 & 0.93684 & IMP & GPN \\
17 & I &
\phantom{1}0.44450 & \phantom{1}4.0781 & \phantom{1}2.8351
 & 1.0224 & IMP & GPN \\
ADP & & \phantom{1}0.29000 & 14.846 & \phantom{1}4.4590 & 1.5360
 & IMP & GPA \\
\hline\hline
\end{tabular}
\caption{Exponents of GPL or DPL density profiles which
minimize the sum of square residuals, using both
analytical (current paper) and numerical  (CMV)
RFSM5 or RFSM4 methods, with regard to 17 sample haloes
listed in Tab.\,\ref{t:SDH}, together with the
averaged density profile (ADP).   Method captions:
GPA - analytical RFSM5; GPN - numerical RFSM5; DPA 
- analytical RFSM4 where GPL density profiles are
reduced to their DPL counterparts i.e. $\alpha\to
+\infty$ and $\chi\to0$.   Other captions as in
Tab.\,\ref{t:numa}.}
\label{t:qmia}
\end{table}
\begin{table}
\begin{tabular}{
lllllcc}
\hline
\hline
\multicolumn{1}{c}{$\xi_{vir}$} &
\multicolumn{1}{c}{$r_0$} &
\multicolumn{1}{c}{$\rho_0$} &
\multicolumn{1}{c}{$\sum\vert R_i\vert$} &
\multicolumn{1}{c}{$\sum R_i^2$} &
\multicolumn{1}{c}{class} &
\multicolumn{1}{c}{method} \\
\hline
8.5258~E$-$2 & 3.2724~E+4 &
1.4808~E$-$7 & 2.7556~E$-$1 & 5.3397~E$-$3 & IMP & GPA \\
9.6963~E+0 & 2.9334~E+2 &
4.8193~E$-$5 & 6.4395~E$-$1 & 3.5625~E$-$2 & BMP & DPA \\
1.9336~E+0 & 1.4710~E+3 &
1.9761~E$-$6 & 9.1288~E$-$1 & 5.1722~E$-$2 & IMP & DPA \\
1.9743~E+0 & 1.4407~E+3 &
2.0941~E$-$6 & 7.0852~E$-$1 & 4.3206~E$-$2 & IMP & DPA \\
7.0536~E+0 & 4.0324~E+2 &
3.0094~E$-$5 & 3.9790~E$-$1 & 1.3080~E$-$2 & EMP & GPA \\
2.0441~E+1 & 1.2999~E+2 &
1.7204~E$-$4 & 3.7559~E$-$1 & 1.2647~E$-$2 & EMP & GPA \\
4.2267~E+0 & 7.4256~E+2 & 1.7070~E$-$4 &
6.1250~E$-$1 & 2.9137~E$-$2 & IMP & GPN \\
5.4741~E$-$1 & 6.4459~E+3 &
1.9954~E$-$4 & 2.4863~E$-$1 & 4.5390~E$-$3 & IMP & GPA \\
9.4211~E+0 & 2.7234~E+2 &
2.0100~E$-$4 & 2.6901~E$-$1 & 5.7462~E$-$3 & EMP & GPA \\
6.6374~E+0 & 3.6395~E+2 & 2.5132~E$-$4 &
5.0765~E$-$1 & 2.0731~E$-$2 & IMP & GPN \\
6.9351~E+0 & 2.8983~E+2 & 3.4633~E$-$5 &
3.3590~E$-$1 & 9.4062~E$-$3 & EMP & GPA \\
1.6379~E+1 & 1.6432~E+2 & 1.2941~E$-$4 &
4.8093~E$-$1 & 2.8344~E$-$2 & EMP & GPA \\
8.0514~E+0 & 2.4964~E+2 & 5.2008~E$-$5 &
6.1515~E$-$1 & 3.2976~E$-$2 & EMP & GPA \\
1.3437~E+0 & 2.0338~E+3 & 3.9969~E$-$5 &
3.8299~E$-$1 & 1.1169~E$-$2 & EMP & GPA \\
8.9712~E+0 & 2.6673~E+2 & 4.6219~E$-$5 &
4.9331~E$-$1 & 2.0331~E$-$2 & EMP & GPA \\
3.7154~E+0 & 5.8599~E+2 & 1.1226~E$-$4 &
4.4819~E$-$1 & 1.5259~E$-$2 & IMP & GPN \\
4.6774~E+0 & 4.9723~E+2 & 1.5144~E$-$4 &
7.1010~E$-$1 & 5.7718~E$-$2 & IMP & GPN \\
2.2073~E+0 & 1.2077~E+3 & 5.9044E$-$6 &
1.1027~E$-$1 & 1.1783~E$-$3 & IMP & GPA \\
\hline\hline
\end{tabular}
\caption{Scale parameters of GPL or DPL density profiles which
minimize the sum of square residuals, using both
analytical (current paper) and numerical (CMV)
RSFM5 or RSFM4 methods, with regard to 17 sample haloes
listed in Tab.\,\ref{t:SDH}, together with the
averaged density profile (ADP).   Method captions:
GPA - analytical RFSM5; GPN - numerical RFSM5; DPA 
- analytical RFSM4 where GPL density profiles are
reduced to their DPL counterparts i.e. $\alpha\to
+\infty$ and $\chi\to0$.   Other captions as in
Tab.\,\ref{t:numa}. Cases and types have been
omitted to save space.}
\label{t:qmib}
\end{table}

The values of some analytical and geometrical
parameters, $\eta_{ADP}$, related to the
best fitting GPL density profile to the mean
SDH density profile, are listed in Tab.\,\ref
{t:parmq} together with
their counterparts, $\overline{\eta}$, averaged
over the best fitting GPL density profiles
to the whole halo sample, using the data of
Tab.\,\ref{t:qmib}.   Also listed in Tab.\,\ref
{t:parmq} are the related standard deviations,
$\sigma_{s~\eta}$, the standard deviations from
the mean, $\sigma_{s~\overline{\eta}}$, and the
standard deviations from the standard
deviation from the mean, $\sigma_{s~\overline
{\mu}}$, which are expressed as (e.g., Oliva \&
Terrasi 1976, Chap.\,V, \S\,5.6.3):
\begin{table}
\begin{tabular}{lrrrrrr}
\hline
\hline
\multicolumn{1}{c}{$\eta$} &
\multicolumn{1}{c}{$(\eta_{ADP})_a$} &
\multicolumn{1}{c}{$(\eta_{ADP})_n$} & 
\multicolumn{1}{c}{$\overline{\eta}$} & 
\multicolumn{1}{c}{$\sigma_{s~\eta}$} &
\multicolumn{1}{c}{$\sigma_{s~\overline{\eta}}$} &
\multicolumn{1}{c}{$\sigma_{s~\overline{\mu}}$} \\
\hline
$\alpha$ & 2.9000~E$-$1 & 5.6832~E$-$1 & 3.0190~E$+$0 &
5.8241~E$+$0 & 1.4125~E$+$0 & 2.4225~E$-$1 \\
$\beta$ & 4.4590~E$+$0 & 3.3143~E$+$0 & 3.8702~E$+$0 &
3.6470~E$+$0 & 8.8452~E$-$1 & 1.5169~E$-$1 \\
$\gamma$ & 1.5360~E$+$0 & 1.0000~E$+$0 & 1.3032~E$+$0 &
3.4209~E$-$1 & 8.2969~E$-$2 & 1.4229~E$-$2 \\
$\xi_{vir}$ & 2.2073~E$+$0 & 3.6308~E$+$0 & 6.5935~E$+$0 &
5.4791~E$+$0 & 1.3289~E$+$0 & 2.2790~E$-$1 \\
$\nu_{mas}$ & 2.1748~E$-$4 & 5.5823~E$-$1 & 1.4501~E$+$1 &
1.8805~E$+$1 & 4.5608~E$+$0 & 7.8217~E$-$1 \\
$l_\xi$ & 3.4386~E$-$1 & 5.6000~E$-$1 & 5.9416~E$-$1 &
5.8669~E$-$1 & 1.4229~E$-$1 & 2.4403~E$-$2 \\
$l_\nu$ & $-$3.6626~E$+$0 & $-$2.5319~E$-$1 &
3.6087~E$-$1 & 1.3833~E$+$0 & 3.3549~E$-$1 & 5.7536~E$-$2 \\
$y_C$ & 7.1609~E$+$0 & 4.4000~E$+$0 & 3.9624~E$+$0 &
8.7244~E$-$1 & 2.1160~E$-$1 & 3.6289~E$-$2 \\
$b$ & 1.8564~E$+$0 & 1.8500~E$+$0 & 1.8410~E$+$0 &
1.0135~E$-$1 & 2.4582~E$-$2 & 4.2157~E$-$3 \\
$b_\beta$ & 5.6277~E$+$0 & 2.5440~E$+$0 & 3.4029~E$+$0 &
4.4866~E$+$0 & 1.0882~E$+$0 & 1.8662~E$-$1 \\
$b_\gamma$ & 7.1081~E$+$0 & 3.8400~E$+$0 & 3.1250~E$+$0 &
7.6132~E$-$1 & 1.8465~E$-$1 & 3.1667~E$-$2 \\
\hline\hline
\end{tabular}
\caption{Comparison between parameters, $\eta
_{ADP}$, related to the best fitting
density profile to the averaged SDH density 
profile (ADP), with regard to analytical
and numerical RFSM5 or RFSM4 methods, respectively,
and their counterparts, $\overline{\eta}$,
averaged over the best fitting density
profiles to the whole halo sample, listed
in Tabs.\,\ref{t:qmia} and \ref{t:qmib}.
Also listed are the related standard deviations,
$\sigma_{s~\eta}$, the standard deviations from
the mean, $\sigma_{s~\overline{\eta}}$, and the
standard deviations from the standard deviation
from the mean, $\sigma_{s~\overline{\mu}}$.
With regard to the exponent, $\alpha$, three
DPL density profiles have been excluded as
$\alpha\to+\infty$ in that limiting situation.
To save space, the notations, $\l_\xi=
\log\xi_{vir}$ and $l_\nu=\log\nu_{mas}$, have
been used.  It is worth remembering that $\log
\xi_{vir}=-x_C$, according to Eq.\,(\ref{eq:xC}).}
\label{t:parmq}
\end{table}
\begin{lefteqnarray}
\label{eq:etam}
&& \overline{\eta}=\frac1n\sum_{i=1}^n\eta_i~~; \\
\label{eq:seta}
&& \sigma_{s{\eta}}=\left[\frac1{n-1}\sum_
{i=1}^n(\eta_i-\overline{\eta})^2\right]^{1/2}~~; \\
\label{eq:setam}
&& \sigma_{s\overline{\eta}}=\left[\frac1n\frac1{n-1}\sum_
{i=1}^n(\eta_i-\overline{\eta})^2\right]^{1/2}~~; \\
\label{eq:smum}
&& \sigma_{s\overline{\mu}}=
\frac{\sigma_{s\overline{\eta}}}{\sqrt{2n}}~~;\quad
\overline{\mu}=\sigma_{s\overline{\eta}}~~;
\end{lefteqnarray}
where $\eta=\alpha$, $\beta$, $\gamma$,
$\xi_{vir}$, $\nu_{mas}$, $\log\xi_{vir}$, 
$\log\nu_{mas}$, $y_C$, $b$, $b_\beta$, $b_\gamma$;
$n=17$ in all cases but $n=14$ if $\eta=\alpha$, as
$\alpha\to+\infty$, related to DPL density profiles,
cannot be considered in averaging;
and, owing to Eq.\,(\ref{eq:xC}), $\log\xi_{vir}=-x_C$.

The following conclusions are deduced from 
Tab.\,\ref{t:parmq}.
\begin{description}
\item[\rm{(x)}\hspace{1.7mm}] Values of 
parameters, $\eta_{ADP}$, related to
the best fitting GPL density profile to
the mean SDH density profile, are different
from their counterparts averaged over the
best fitting GPL density profiles to the
whole halo sample, listed in Tabs.\,\ref
{t:qmia} and \ref{t:qmib}.   The discrepancy
changes for different parameters, from
less than $\mp\sigma_{s~\bar{\eta}}$ to
more than $\mp3\sigma_{s~\bar{\eta}}$.
The above result is expected from the
theory of the errors.
\item[\rm{(xi)}] The exponents of best
fitting, GPL density profiles related
to analytical RSFM5 methods, are distant
from their NFW counterparts, $(\alpha,
\beta,\gamma)=(1,3,1)$, conform to
$[\Nint(\alpha),\Nint(\beta),\Nint(\gamma)]
=(0,4,2)$, where $\Nint$ denotes the
nearest integer.   The difference increases
from about one half for $\gamma$
to about one and half for $\beta$.
On the other hand, the contrary holds
with regard to GPL density profiles
related to numerical RFSM5 methods,
namely $[\Nint(\alpha),\Nint(\beta),
\Nint(\gamma)]=(1,3,1)$, where the
difference increases from about one
hundredth for $\gamma$ to about one
half for $\alpha$ (CMV).
\item[\rm{(xii)}\hspace{0.0mm}] The rsm
error of the logarithm of the scaled radius,
$\xi_{vir}$, is $\sigma_{s~\log\xi_{vir}}
\approx0.60$, to be compared with $\sigma_{s~
\log\xi_{vir}}=0.18$ deduced from richer
samples where $(\alpha,\beta,\gamma)=(1,3,1)$;
$M_{vir}=$(0.5-1.0)$\times10^nh^{-1}{\rm m}_
\odot$; $11\le n\le14$; and $n$ is an integer
(Bullock et al. 2001).
\end{description}

\subsection{Discussion}\label{disc}

The application of an analytical RSFM5 method
makes a necessary condition for the occurrence
of an extremal minimum, Eqs.\,(\ref{seq:gay}),
which is satisfied in 8 sample haloes on a total
of 17, listed in Tab.\,\ref{t:SDH}.   The
related points in the 5-dimension parameter
hyperspace, ${\sf P}^\ast\equiv(x_C^\ast,y_
C^\ast,\alpha^\ast,\beta^\ast,\gamma^\ast)$,
may safely be thought of as absolute extremal
minima, by comparison with fiducial minima
calculated using a numerical RSFM5 method
(CMV).   On the other hand, the existence
of other extremal minimum points cannot be
excluded in absence of a rigorous mathematical
proof.   In addition, the lack of extremal
minima in the remaining 9 sample haloes plus
ADP, could be an artifact related to the
algorithm used for determining the intersection
of the three surfaces, $\gamma=\gamma(x_C,
\alpha)$, expressed by Eqs.\,(\ref{seq:gay}).
An improved algorithm would suceed in
localizing extremal minimum points in a
larger number of sample haloes.

The occurrence of an extremal minimum, or
the algorithm power in finding it, seems
to be correlated with halo morphology.
The relative frequency of the above mentioned
event is 7/12$\approx$0.583 for safely
relaxed haloes, and 1/5=0.2 for safely
undergoing a major merger haloes.

The presence of a (safely absolute)
extremal point of minimum, makes a firm
criterion for deciding which, among the
family of GPL density profiles, best fits
an assigned SDH density profile.   In
absence of extremal minima, numerical
and analytical RFSM5 or RFSM4 methods,
the last using DPL instead of GPL density
profiles, seem to be competitive.   With
regard to sample haloes where no extremal
minimum point has been found, the sum of
square residuals is minimized using a
numerical RFSM5 method in 3 cases, an
analytical RFSM5 method in 2 cases,
and an analytical RFSM4 method in 3
cases, as can be seen in Tab.\,\ref
{t:srq}.

The related parameters, including cases
where an extremal point of minimum
occurs, together with ADP, are shown in
Tabs.\,\ref{t:qmia} and \ref{t:qmib}.
The exponents, $\alpha$, $\beta$,
$\gamma$, range within $0.42<\alpha<
+\infty$ ($\alpha=0.29$ for ADP); $0<
\chi<13$ ($\chi=14.846$ for ADP); $2.0
<\beta<17.6$; $0.6<\gamma<1.9$, respectively.
The scaled radius, $\xi_{vir}$, the scaling
radius, $r_0$, and the scaling density,
$\rho_0$, range within $0.08<\xi_{vir}<
20.5$; $129<r_0/{\rm kpc}<32724$; $0.001<
10^4\rho_0/(10^{10}{\rm m}_\odot/{\rm
kpc}^3)<2.6$; respectively.   The range
in inner slope, $\gamma$, is consistent
with both core-shaped ($\gamma<1$) and
cuspy ($\gamma\ge1$) density profiles
\footnote{Strictly speaking, a central
cusp occurs for $\gamma>0$.}.
The range in outer slope, $\beta$, is
consistent with both infinite ($\beta
\le3$) and finite ($\beta>3$) total
(with no truncation) mass.

With regard to the exponent, $\gamma$,
it is worth mentioning that, under reasonable
boundary conditions such as force-free
halo centre and vanishing density at
infinite distance, the inequality
$\gamma<1$ holds (M\"ucket \& Hoeft 2003).
On the other hand, simple analytic
considerations suggest that $\gamma\le2$
(Williams et al. 2004).  Finally, the
validity of Jeans equation implies
$1\le\gamma\le3$ (Hansen 2004), where
shallower slopes can arise, provided
the effects of baryonic matter are
considered (El-Zant et al. 2004; Hansen 2004).
The current result, $0.6\le\gamma\le1.9$,
is consistent, or marginally consistent,
with the above mentioned constraints.

\begin{figure}
\centerline{\psfig{file=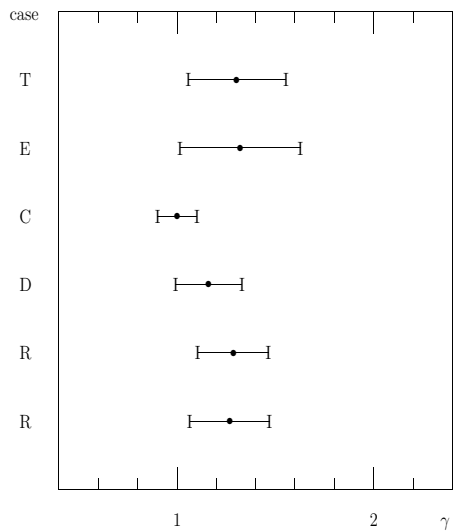,height=180mm,width=110mm}}
\caption{Mean values and related scatters, $\overline
{\gamma}\mp3\sigma_{s~\overline{\gamma}}$, corresponding
to cases listed in Tab.\,\ref{t:gamma}, positioned and
labeled as therein.}
\label{f:inte}    
\end{figure}

On the other hand, values of asymptotic
inner slopes of fitting density profiles
determined in the current
paper, are consistent with their counterparts
deduced from recent high-resolution
simulations using a three-parameter fit
involving scaling radius, scaling density,
and asymptotic inner slope (6 sample objects,
Diemand et al. 2004) or a two-parameter fit
involving scaling radius and asymptotic inner
slope (16 sample objects, Reed et al. 2004).
The related parameters are listed in Tab.\,\ref
{t:gamma}, which shows agreement between
different approaches, within the fiducial range,
$\overline{\gamma}\mp3\sigma_{s~\overline{\gamma}}$.
\begin{table}
\begin{tabular}{clllllllll}
\hline
\hline
\multicolumn{1}{c}{c} &
\multicolumn{1}{c}{$n$} &
\multicolumn{1}{c}{$\overline{\gamma}$} &
\multicolumn{1}{c}{$\sigma_{s~\gamma}$} &
\multicolumn{1}{c}{$\sigma_{s~\overline{\gamma}}$} &
\multicolumn{1}{c}{$\Delta^-\gamma$}  &
\multicolumn{1}{c}{$\Delta^+\gamma$} &
\multicolumn{1}{c}{$\Delta^\mp\gamma$} &
\multicolumn{1}{c}{$\overline{\gamma}^-$} &
\multicolumn{1}{c}{$\overline{\gamma}^+$} \\

\hline
T &            17  & 1.3032  & 0.34209 & 0.082969 & 0.70036 & 0.43350 & 0.56693 & 1.0543  & 1.5521 \\
E & $\phantom{1}8$ & 1.3218  & 0.28798 & 0.10181  & 0.45124 & 0.38076 & 0.41600 & 1.0164  & 1.6273 \\
C &            17  & 1.0018  & 0.13893 & 0.033696 & 0.29868 & 0.19820 & 0.24844 & 0.90071 & 1.1029 \\
D & $\phantom{1}6$ & 1.1617  & 0.13732 & 0.056060 & 0.24167 & 0.25833 & 0.25000 & 0.99352 & 1.3299 \\
R &            16  & 1.2875  & 0.23910 & 0.059774 & 0.28750 & 0.41250 & 0.35000 & 1.1082  & 1.4668 \\
R &            13  & 1.2692  & 0.24285 & 0.067353 & 0.26923 & 0.43077 & 0.35000 & 1.0671  & 1.4713 \\
\hline\hline
\end{tabular}
\caption{Comparison between statistical parameters related
to the asymptotic inner slope, $\gamma$, deduced from
different samples using different fits.   Column captions:
1 - case: T - current paper, all sample haloes; E - current
paper, only sample haloes where an extremum point of minimum
occurs; C - Caimmi et al. (2004); D - Diemand et al. (2004);
R - Reed et al. (2004); 2 - $n$: total number of sample
objects; 3 - $\overline{\gamma}$: arithmetic mean; 4 -
$\sigma_{s~\gamma}$: standard deviation; 5 - $\sigma_
{s~\overline{\gamma}}$: standard deviation from the mean;
6 - $\Delta^-\gamma$: maximum negative deviation from
the mean; 7 - $\Delta^+\gamma$: maximum positive deviation
from the mean; 8 - $\Delta^\mp\gamma$: mean maximum deviation
from the mean; 9 - $\overline{\gamma}^-=\overline{\gamma}-3
\sigma_{s~\overline{\gamma}}$: lower limit assigned to the
mean; 10 - $\overline{\gamma}^+=\overline{\gamma}+
3\sigma_{s~\overline{\gamma}}$: upper limit assigned to the
mean.   Sample haloes represent clusters or groups, with
the exception of three objects belonging to the richer
R sample (excluded in the poorer R sample), which
represent SDHs embedding the Milky Way and two dwarf
galaxies, respectively.}
\label{t:gamma}
\end{table}
A marginal discrepancy between results from CMV
and Reed et al. (2004) can be eliminated, by
removing from the latter sample three objects, which
represent SDHs embedding the Milky Way and two dwarf
galaxies, respectively, instead of clusters or groups.
On the other hand, current results make an improvement
with respect to CMV.
Mean values and related scatters, $\overline{\gamma}\mp
3\sigma_{s~\overline {\gamma}}$, corresponding to each
case of Tab.\,\ref{t:gamma}, are represented in
Fig.\,\ref{f:inte}.

No evident correlation is found between SDH 
dynamical state (relaxed or merging) and
asymptotic inner slope of the logarithmic
density profile, $-\gamma$, or (for SDH comparable
virial masses) scaled radius, $\xi_{vir}$,
contrary to previous results (Ascasibar et al.
2004) related to a sample of 19 high-resolution
SDHs on the scale of both clusters of galaxies
(13 objects) and galaxies (6 objects), with regard 
to NFW density profiles.   An investigation on
richer samples could provide more information
to this respect.

The large range in parameters shown in
Tabs.\,\ref{t:qmia} and \ref{t:qmib},
provides further evidence to the
occurrence of a certain degree of
degeneracy in fitting GPL to SDH density
profiles (Klypin et al., 2001; CMV).
For instance, with regard to case 14
listed in Tabs.\,\ref{t:numa} and \ref
{t:numb}, using analytical and numerical
RFSM5 methods, the sum of square residuals
changes by less than 10\%, while the
remaining parameters change by more
than 40\%.

As different methods are involved, it
seems difficult that the occurrence of
degeneracy is intrinsic to the procedure
of minimization.   A remaining possibility
could be the lack of resolution in the
central regions, which makes the lower
limit of the domain, defined by Eq.\,(\ref
{eq:doeta}).   A more extended (on the left)
range, owing to the next generation of
high-resolution simulations, could reduce
or eliminate the degeneracy.   The fit
has necessarily to be restricted to the
virialized region of the halo, which
leaves the upper limit of the domain
unchanged.

The occurrence of degeracy could also
bias the procedure used for minimizing
the sum of any function (in particular,
the square) of residuals.   For instance,
the application of a numerical RFSM5
method to the halo sample listed in
Tab.\,\ref{t:SDH} yields an inner slope,
$\gamma\approx1$ (CMV), different from
a value $\gamma\approx1.5$ currently
found for dark matter haloes on the
scale of cluster of galaxies (e.g.,
Fukushige \& Makino 2003; Hiotelis
2003), but closer to $\gamma\approx
1.2$-1.3 from more recent attempts
(Diemand et al. 2004; Reed et al. 
2004).   On the other hand, the
application of an analytical RFSM5
method to the above mentioned halo
sample yields an inner slope, $\gamma
\approx1.5$ (current paper).

The scaled radius, $\xi_{vir}=r_{vir}/r_0$,
is defined in the current paper as related
to the point where the maximum change in
logarithmic slope occurs in the corresponding
GPL density profile (CMV).   On the other
hand, the concentration, $c=r_{vir}/r_{-2}$, 
is usually defined as related to the point
where the logarithmic slope equals $-2$ (e.g.,
Klypin et al. 2001; Bullock et al. 2001;
Hiotelis 2003).   The former definition
seems to be more general, as it provides
the scaled radius meaningful also in early
times, where the slope of a GPL density
profile may be smaller (in absolute value)
than 2, while the contrary holds for the
concentration (Hiotelis 2003).

The GPL density profile which best fits
to the averaged SDH density profile,
with regard to analytical RFSM5 methods, is
characterized by exponents, $(\alpha,\beta,
\gamma)$, satisfying $[\Nint(\alpha),\Nint
(\beta),\Nint(\gamma)]=$(0,4,2), which is
different from NFW density profiles
currently used in literature, $(\alpha,\beta,
\gamma)=(1,3,1)$.   The comparison with
values averaged over the whole halo sample,
discloses that fluctuations of the exponents
around their NFW counterparts may safely be
excluded.   Accordingly, NFW density profiles
cannot be conceived as intrinsic to dark
matter haloes of the current sample.

On the other hand, following e.g., Bullock
et al. (2001), NFW density profiles (or
alternative functional forms) may be
considered as a convenient way to parametrize
SDH density profiles, without implying that
it necessarily provides the best possible fit.
This is why the scaled radius, $\xi_{vir}$,
and the scaled mass, $\nu_{mas}$, can be
interpreted as general structure parameters,
not necessarily restricted to a specific
density profile.   In particular, any spread 
in $\xi_{vir}$ and $\nu_{mas}$ can be
attributed to a real (larger) scatter in a
``physical'' scaling radius, defined by e.g.,
the radius where the change in slope of the
logarithmic density profile attains its
maximum value (CMV), rather than to
inaccuracies in an assumed, ``universal''
density profile.   For further details see
e.g., Bullock et al. (2001).

Additional support to the above considerations
is provided by the value calculated for the
standard deviation of the decimal logarithm
of the scaled radius, $\sigma_{s\log\xi_{vir}}
\approx0.59$, which is 3.3-3.4 times larger
than $\sigma_{s\log\xi_{vir}}\approx0.18$
deduced from a statistical
sample of about five thousands simulated
haloes, within mass bins equal to (0.5-1.0)
$\times10^nh^{-1}{\rm m}_\odot$, where $11\le
n\le14$ and $n$ is an integer, with regard to
NFW density profiles (Bullock et al. 2001).
A comparable scatter, $\sigma_{s\log\xi_{vir}}
\approx0.17$, has been obtained using numerical
RSFM5 methods (CMV).   The discrepancy with
the results of the current paper, is probably
owing to a bias in the selection of the
subdomain where numerical RFSM5 methods work.

\begin{figure}
\centerline{\psfig{file=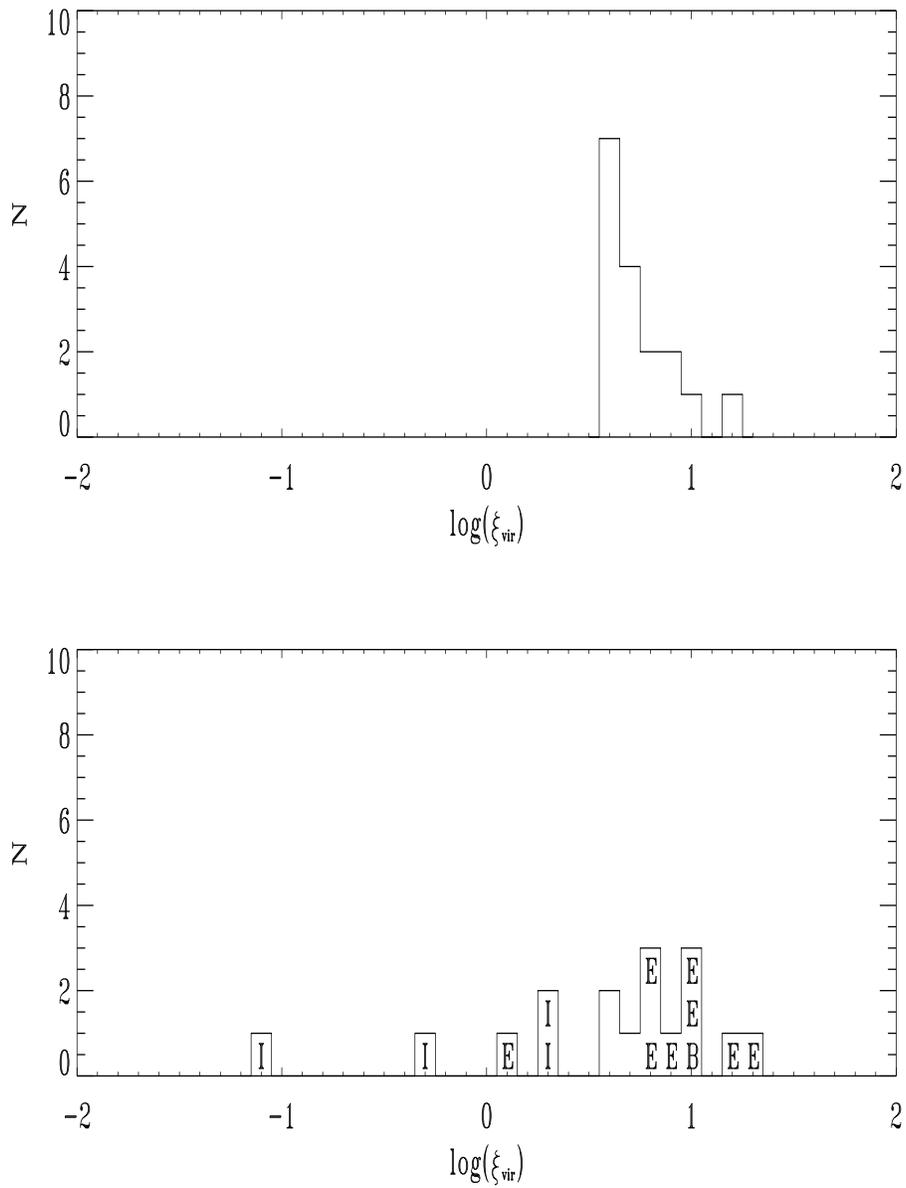,height=170mm,width=130mm}}
\caption{Histogram of the decimal logarithm of the
scaled radius, $\log\xi_{vir}$, with regard to
numerical (CMV; top) and best fitting (current paper, 
Tab.\,\ref{t:qmib}; bottom) RSFM5 methods.
Label captions: I-IMP; E-EMP; B-BMP; blank -
numerical (CMV) RSFM5 methods i.e. same cases as
in the upper istogram.}
\label{f:isto}    
\end{figure}

To get further insight, the histogram of the
decimal logarithm of the scaled radius, $\log
\xi_{vir}$, is represented in Fig.\,\ref{f:isto},
with regard to numerical (CMV) and best fitting
(current paper, Tab.\,\ref{t:qmib}) RSFM5 methods.
The former case (top panel) shows a clear cut at
$\log\xi_{vir}=0.55$, which could be a biasing
effect, as mentioned above, but the dispersion
is low.   The latter case (bottom panel) shows
no sign of cut, but the dispersion is large.
It is reduced only slightly by restricting the
calculations to a homogeneous subsample, where
only EMPs are considered, and even if the
maximum deviation from the arithmetic mean is
excluded therein.   If a large dispersion still
maintains for richer samples $(N\gg17)$, in
dealing with analytical RSFM5 methods, a low
dispersion found in $N$-body simulations (e.g.,
Bullock et al. 2001) seems to
be an artefact, related to the assumption of NFW
(or any equivalent choice) density profile.

On the other hand, a large spread in parameters
of GPL density profiles where an extremum point
of minimum occurs, may be considered as fiducial
at most.   This is in absence of a rigorous
proof on the existence of an absolute point of
minimum.   If, in fact, absolute minima do not
coincide with the extremum points of minimum
which are found in the
current attempt, then the spread in parameters
could be reduced.

The results of the current paper confirm a
certain degree of degeneracy in fittig GPL
to SDH density profiles, as pointed out by
Klypin et al. (2001).   For instance, cases
4 and 14 of Tabs.\,\ref{t:numa} and \ref
{t:numb} exhibit - on the related upper and
lower lines - values of the sum of square
residuals and absolute value residuals,
which are close each to the other.   On the
contrary, the difference substantially grows
for the remaining parameters.   As degeneracy
occurs using analytical and numerical RFSM5
methods separately, it cannot safely be
considered as an artifact of the corresponding
algorithm.   It might be either a real
feature of the problem of minimization,
or a consequence of an incomplete domain,
towards central regions, of SDH density
profiles, owing to numerical artifacts
(mainly two-body relaxation).   The latter
possibility could be tested by the next
generation of high-resolution simulations.

The current attempt is limited to GPL
density profiles, defined by Eq.\,(\ref
{eq:GPL}), but different alternatives
have been exploited in the literature,
such as the family of density profiles:
\begin{equation}
\label{eq:prob}
\rho\left(\frac r{r_0}\right)=\rho_0\left[
1+\left(\frac r{r_0}\right)^
\alpha\right]^{-\chi}\left[1+\frac r{r_0}
\right]^{-\gamma}~~;\qquad\chi=\frac{\beta-
\gamma}\alpha~~;
\end{equation}
where the choice $(\alpha,\beta,\gamma)=
(2,3,1)$ corresponds to the Burkert (1995)
density profile, which resembles the NFW
density profile for $r\appgeq0.02r_{vir}$
(e.g. Bullock et al. 2001).   

Another possibility is a profile that curves
smoothly over to a constant density at very
small radii (Navarro et al. 2004):
\begin{equation}
\label{eq:expro}
\rho=\rho_0\exp\left\{-\frac2\lambda\left[
\left(\frac rr_0\right)^\lambda-1\right]
\right\}~~;
\end{equation}
where the parameter, $\lambda$, prescribes
how fast the density profile turns away from
a power-law near the centre.   The best fit
reads (19 sample objects, Navarro et al. 2004):
$\overline{\lambda}\mp3\sigma_{s~\overline
{\lambda}}=0.17216\mp0.021897$.

On the other hand, the RFSM5 method may be
extended to any kind of density profile.

\section{Conclusion}\label{conc}

Analytical and geometrical properties
of generalized power-law (GPL) density profiles
have been reviewed, and special effort has been
devoted to the special cases where GPL density
profiles reduce to (i) a double power-law
(DPL), and (ii) a single power-law (SPL).
Then GPL density profiles have been compared with
simulated dark haloes (SDH) density profiles,
and nonlinear least-squares fits have been prescribed,
involving five parameters (a scaling radius,
$r_0$, a scaling density, $\rho_0$, and three
exponents, $\alpha$, $\beta$, $\gamma$), which
specify the fitting profile (RFSM5 method).

More specifically, the validity of a necessary
condition for the occurrence of an extremum
point, has been related to the existence of an
intersection between three surfaces, $\gamma_k=
f_k(r_0,\alpha)$, $k=1,2,3,$ in a 3-dimension
space, $({\sf O}~r_0~\alpha~\gamma)$.   Using 
the algorithm has
made also establish that the extremum point
is a fiducial minimum, while the explicit
calculation of the Hessian determinant has been
avoided to gain in simplicity.   In absence
of a rigorous proof, the fiducial minimum
could be considered as nothing but a fiducial 
absolute minimum.

An application has been
made to a sample of 17 SDHs on the scale of
cluster of galaxies, within a flat $\Lambda$CDM
cosmological model (Rasia et al. 2004).   In dealing
with the averaged SDH density profile (ADP), a virial
radius, $r_{vir}$, equal to the mean over the whole
sample, has been assigned, which has allowed the
calculation of the remaining parameters.   The following
results have been found.   (i) A necessary condition
for the occurrence of an extremum point is
satisfied for 8 sample haloes, and is not for
the remaining 9 together with ADP.    In the
former alternative, an extremum point of minimum
(EMP) may safely exist.   In the latter alternative,
the occurrence of an EMP cannot be excluded,
but only a local minimum can be determined.
(ii) The occurrence of an EMP implies a sum of
square residuals which is systematically lower
than its counterpart deduced by use of numerical
RFSM5 methods.   Accordingly, EMPs may safely be
thought of as absolute minima.   With regard to
sample haloes where no EMP is detected, the above
result maintains in 4 cases (including ADP),
while the contrary holds for the remaining 5
cases.   (iii) The best fit (with no EMP detected)
is provided by DPL density profiles for 3 sample
haloes.   In addition, DPL density profiles make
a rough, but viable approximation in fitting SDH
density profiles.   The contrary holds for SPL
density profiles.

No evident correlation has been found between SDH 
dynamical state (relaxed or merging) and
asymptotic inner slope of the logarithmic
density profile or (for SDH comparable
virial masses) scaled radius.

Mean values and standard deviations of some
parameters have been calculated 
and, in particular, the decimal logarithm of the
scaled radius, $\xi_{vir}$, has
read $<\log\xi_{vir}>\,\approx0.59$ and $\sigma_{s
\log\xi_{vir}}\approx0.59$, the standard deviation
exceeding by a factor 3.3-3.4 its counterpart
evaluated in an earlier attempt using NFW
density profiles (Bullock et al. 2001).
If a large dispersion still maintains for richer
samples, in dealing with analytical RSFM5 methods,
a low dispersion found in $N$-body simulations
has seemed to be an artefact,
due to the assumption of NFW (or any equivalent choice)
density profile.
It has provided additional support to the idea, that
NFW density profiles may be considered as a
convenient way to parametrize SDH density
profiles, without implying that it necessarily
produces the best possible fit (Bullock et al. 2001).

With regard to RFSM5 methods formulated
in the current paper, the exponents of both the
best fitting GPL density profile to ADP, $(\alpha,
\beta,\gamma)\approx(0.3,4.5,1.5)$, and related
averages calculated over the whole halo sample,
$(\overline{\alpha},\overline{\beta},\overline
{\gamma})\approx(3.0,3.9,1.3)$, have been found
to be far from
their NFW counterparts, $(\alpha,\beta,\gamma)
=(1,3,1)$.   The last result, together with
the large value of the standard deviation,
$\sigma_{s\log\xi_{vir}}$, has been interpreted as
due to a certain degree of degeneracy in
fitting GPL to SDH density profiles.
If it is a real feature of the problem, or it 
could be reduced by the next generation of
high-resolution simulations, still has remained
an open question.  Values of asymptotic inner slope
of fitting logarithmic density profiles, have
been found to be
consistent with results from recent high-resolution
simulations (Diemand et al. 2004; Reed et al. 2004).

\section{Acknowledgements}
We are endebted with  E. Rasia, G. Tormen, and
L. Moscardini, for making the results of their
simulations available to us.   In addition, we
are deeply grateful to all of them for clarifying
and fruitful discussions.

\appendix

\section{On the occurrence of extremum points of minimum to
the sum of square residuals}\label{EMP}

A necessary condition for the occurrence of
extremum points of minimum to the sum of square residuals,
defined by Eq.\,(\ref{eq:sRiq}), is expressed
by Eqs.\,(\ref{eq:sd50}), or more explicitly as:
\begin{leftsubeqnarray}
\slabel{eq:msRa}
&& \sum_{i=1}^nR_i\frac{\partial R_i}{\partial x_C}=0~~; \\
\slabel{eq:msRb}
&& \sum_{i=1}^nR_i\frac{\partial R_i}{\partial y_C}=0~~; \\
\slabel{eq:msRc}
&& \sum_{i=1}^nR_i\frac{\partial R_i}{\partial\alpha}=0~~; \\
\slabel{eq:msRd}
&& \sum_{i=1}^nR_i\frac{\partial R_i}{\partial\beta}=0~~; \\
\slabel{eq:msRe}
&& \sum_{i=1}^nR_i\frac{\partial R_i}{\partial\gamma}=0~~;
\label{seq:msR}
\end{leftsubeqnarray}
where, for sake of simplicity, it is 
intended that the residuals and their 
first derivatives are calculated at a
fiducial extremum point of minimum, ${\sf P}
\equiv(x_C^\ast,y_C^\ast,\alpha^\ast,
\beta^\ast,\gamma^\ast)$.

By use of Eq.\,(\ref{eq:Ri}), the first
derivatives of the $i$-th residual read:
\begin{leftsubeqnarray}
\slabel{eq:dpRa}
&& \frac{\partial R_i}{\partial x_C}=
-\gamma-(\beta-\gamma)t_i~~; \\
\slabel{eq:dpRb}
&& \frac{\partial R_i}{\partial y_C}=
-1~~; \\
\slabel{eq:dpRc}
&& \frac{\partial R_i}{\partial\alpha}=
-\frac{\beta-\gamma}{\alpha^2}s_i+\frac
{\beta-\gamma}\alpha(x_i-x_C)t_i~~; \\
\slabel{eq:dpRd}
&& \frac{\partial R_i}{\partial\beta}=
\frac1\alpha s_i~~; \\
\slabel{eq:dpRe}
&& \frac{\partial R_i}{\partial\gamma}=
(x_i-x_C)-\frac1\alpha s_i~~;
\label{seq:dpR}
\end{leftsubeqnarray}
where both $s_i$ and $t_i$ depend on
$w_i=\exp_{10}[\alpha(x_i-x_C)]$,
according to Eqs.\,(\ref{eq:si}) and
(\ref{eq:ti}), respectively.   In
addition, the validity of the relations:
\begin{lefteqnarray}
\label{eq:sgb}
&& \frac{\partial R_i}{\partial\gamma}+
\frac{\partial R_i}{\partial\beta}=
(x_i-x_C)~~; \\
\label{eq:sabx}
&& \frac{\partial R_i}{\partial\alpha}+
\frac{x_i-x_C}\alpha
\frac{\partial R_i}{\partial x_C}+
\frac{\beta-\gamma}\alpha
\frac{\partial R_i}{\partial\beta}=
-\frac\gamma\alpha(x_i-x_C)~~;
\end{lefteqnarray}
may easily be checked.

The combination of Eqs.\,(\ref{eq:msRb}) and
(\ref{eq:dpRb}) yields:
\begin{equation}
\label{eq:sR0}
\sum_{i=1}^nR_i=0~~;
\end{equation}
and the combination of Eqs.\,(\ref{eq:msRd}),
(\ref{eq:msRe}), (\ref{eq:sgb}), and
(\ref{eq:sR0}), yields:
\begin{equation}
\label{eq:sRx0}
\sum_{i=1}^nR_ix_i=0~~;
\end{equation}
on the other hand, multiplying the left
and the right-hand side of Eq.\,(\ref
{eq:sabx}) by $R_i$ and summing over
all the residuals, produces:
\begin{lefteqnarray}
\label{eq:ssabx}
&& \sum_{i=1}^nR_i\frac{\partial R_i}
{\partial\alpha}+\sum_{i=1}^n\frac
{x_i-x_C}\alpha R_i\frac{\partial R_i}
{\partial x_C} \nonumber \\
&& +\frac{\beta-\gamma}\alpha\sum_{i=1}^n
R_i\frac{\partial R_i}{\partial\beta}=-\frac
\gamma\alpha\sum_{i=1}^nR_i(x_i-x_C)~~; 
\end{lefteqnarray}
which, owing to Eqs.\,(\ref{eq:msRa}),
(\ref{eq:msRc}),
(\ref{eq:msRd}), (\ref{eq:sR0}), and
(\ref{eq:sRx0}), yields:
\begin{equation}
\label{eq:xRd0}
\sum_{i=1}^nx_iR_i\frac{\partial R_i}
{\partial x_C}=0~~;
\end{equation}
similarly, the combination of Eqs.\,(\ref
{eq:msRa}), (\ref{eq:dpRa}), (\ref{eq:sR0}),
and (\ref{eq:xRd0}), yields:
\begin{equation}
\label{eq:sRt0}
\sum_{i=1}^nR_it_i=0~~;
\end{equation}
and the combination of Eqs.\,(\ref
{eq:msRd}) and (\ref{eq:dpRd}) yields:
\begin{equation}
\label{eq:sRs0}
\sum_{i=1}^nR_is_i=0~~;
\end{equation}
finally, the combination of Eqs.\,(\ref
{eq:msRc}), (\ref{eq:dpRc}), (\ref{eq:sRt0}),
and (\ref{eq:sRs0}), yields:
\begin{equation}
\label{eq:sRxt0}
\sum_{i=1}^nR_ix_it_i=0~~;
\end{equation}
in conclusion, Eqs.\,(\ref{eq:sR0}),
(\ref{eq:sRx0}), (\ref{eq:sRt0}),
(\ref{eq:sRs0}), and (\ref{eq:sRxt0}),
make a system of five independent
equations which is equivalent to
(\ref{seq:msR}).

The substitution of Eqs.\,(\ref{eq:Ri})
and (\ref{eq:si}) into (\ref{eq:sR0})
yields:
\begin{equation}
\label{eq:eR0}
(\overline{y}-y_C)+\gamma(\overline{x}-
x_C)+\frac{\beta-\gamma}\alpha\overline
{s}=0~~;
\end{equation}
where overbars denote mean values, 
according to Eq.\,(\ref{eq:bsb}).
The comparison between alternative
expressions of $y_C$, deduced from
Eqs.\,(\ref{eq:Ri}) and (\ref{eq:eR0})
produces:
\begin{equation}
\label{eq:R20}
R_i=(y_i-\overline{y})+\gamma(x_i-\overline
{x})+\frac{\beta-\gamma}\alpha(s_i-\overline
{s})~~;
\end{equation}
where no dependence on $y_C$ occurs.

The substitution of Eq.\,(\ref{eq:R20})
into (\ref{eq:sRx0}) yields:
\begin{equation}
\label{eq:Rx20}
\upsilon_s(x,y)+\gamma\upsilon_s(x,x)+
\frac{\beta-\gamma}\alpha\upsilon_s(x,
s)=0~~;
\end{equation}
where the function, $\upsilon_s$, is
defined by Eqs.\,(\ref{seq:bs}), and
can be conceived as a two-dimension
(2D) empirical variance.   This is
why, in the limit of a one-dimension
(1D) empirical variance, $\upsilon_s
(a,a)=\overline{(a^2)}-(\overline{a})^2$,
according to the standard notation.
It is worth of note that, while 1D
empirical variances are always non
negative, the contrary holds for 2D
empirical variances.

The comparison between alternative
expressions of $\chi=(\beta-\gamma)/
\alpha$, deduced from Eqs.\,(\ref
{eq:R20}) and (\ref{eq:Rx20}),
produces:
\begin{lefteqnarray}
\label{eq:R30}
&& R_i=(y_i-\overline{y})+\gamma(x_i-
\overline{x}) \nonumber \\
&& \phantom{R_i=}-\frac{\upsilon_s(x,y)-
\gamma\upsilon_s(x,x)}{\upsilon_s(x,s)}
(s_i-\overline{s})~~;
\end{lefteqnarray}
where no dependence on $\beta$
occurs.   Then the combination of
Eqs.\,(\ref{eq:sRt0}) and (\ref{eq:R30})
yields:
\begin{lefteqnarray}
\label{eq:Rt20}
&& \upsilon_s(x,s)\upsilon_s(y,t)-
\upsilon_s(s,t)\upsilon_s(x,y)= \nonumber \\
&& \gamma[\upsilon_s(s,t)\upsilon_s(x,x)-
\upsilon_s(x,s)\upsilon_s(x,t)]~~;
\end{lefteqnarray}
and the combination of Eqs.\,(\ref{eq:R30})
and (\ref{eq:Rt20}) allows an expression of 
$R_i$ which depends only on $x_C$ and 
$\alpha$, via
Eqs.\,(\ref{eq:si}) and (\ref{eq:ti}).

The combination of Eqs.\,(\ref{eq:sRs0})
and (\ref{eq:R30}) yields:
\begin{lefteqnarray}
\label{eq:Rs20}
&& \upsilon_s(x,s)\upsilon_s(y,s)-
\upsilon_s(s,s)\upsilon_s(x,y)= \nonumber \\
&& \gamma[\upsilon_s(s,s)\upsilon_s(x,x)-
\upsilon_s(x,s)\upsilon_s(x,s)]~~;
\end{lefteqnarray}
which has the same formal expression as
Eq.\,(\ref{eq:Rt20}), provided the following
correspondences hold: $(y,t)\leftrightarrow
(y,s)$; $(s,t)\leftrightarrow(s,s)$; $(x,t)
\leftrightarrow(x,s)$.

The combination of Eqs.\,(\ref{eq:sRxt0})
and (\ref{eq:R30}) yields:
\begin{lefteqnarray}
\label{eq:Rxt20}
&& \upsilon_s(x,s)\upsilon_s(y,xt)-
\upsilon_s(s,xt)\upsilon_s(x,y)= \nonumber \\
&& \gamma[\upsilon_s(x,x)\upsilon_s(s,xt)-
\upsilon_s(x,s)\upsilon_s(x,xt)]~~;
\end{lefteqnarray}
which has the same formal expression as
Eqs.\,(\ref{eq:Rt20}), and (\ref{eq:Rs20}),
provided the following
correspondences hold: $(y,t)\leftrightarrow
(y,s)\leftrightarrow(y,xt)$; $(s,t)
\leftrightarrow(s,s)\leftrightarrow(s,xt)$;
$(x,t)\leftrightarrow(x,s)\leftrightarrow
(x,xt)$.

Then we are left with three alternative
expressions of $\gamma$, which may be
deduced from Eqs.\,(\ref{eq:Rt20}), (\ref
{eq:Rs20}), (\ref{eq:Rxt20}), and are
expressed by Eqs.\,(\ref{eq:gayt}), (\ref
{eq:gays}), (\ref{eq:gayxt}), respectively.

A necessary condition for the occurrence
of extremum points of  minimum to the sum of square
residuals, defined by Eq.\,(\ref{eq:sRiq}),
has been formulated via Eqs.\,(\ref
{seq:msR}).   The condition is also
sufficient, provided the Hessian
determinant satisfies Eq.\,(\ref{eq:Hesp}).
The derivation of both sides of Eq.\,(\ref
{eq:sd50}) yields:
\begin{leftsubeqnarray}
\slabel{eq:sd2p}
&& \left(\frac{\partial^2 F}{\partial u_j^2}
\right)_{{\sf P}^\ast}=2\sum_{i=1}^n\left[
\left(\frac{\partial R_i}{\partial u_j}
\right)_{{\sf P}^\ast}^2+R_i
\left(\frac{\partial^2R_i}{\partial u_j^2}
\right)_{{\sf P}^\ast}\right]~~; \\
\slabel{eq:sd2m}
&& \left(\frac{\partial^2 F}{\partial u_ju_k}
\right)_{{\sf P}^\ast}=2\sum_{i=1}^n\left[
\left(\frac{\partial R_i}{\partial u_j}
\right)_{{\sf P}^\ast}
\left(\frac{\partial R_i}{\partial u_k}
\right)_{{\sf P}^\ast}+R_i
\left(\frac{\partial^2R_i}{\partial u_j
\partial u_k}\right)_{{\sf P}^\ast}\right]~~;
\label{seq:sd2}
\end{leftsubeqnarray}
where $u_1=x_C$, $u_2=y_C$, $u_3=
\alpha$, $u_4=\beta$, $u_5=\gamma$.

The second derivatives of the $i$-th
residuals may be determined, using
Eqs.\,(\ref{eq:si}), (\ref{eq:ti}),
(\ref{eq:wi}), (\ref{eq:dwa}), (\ref
{eq:dwx}), and (\ref{seq:dpR}).   The
result is:
\begin{leftsubeqnarray}
\slabel{eq:dRxx}
&& \frac{\partial^2 R_i}{\partial x_C^2}=
\alpha(\beta-\gamma)\frac{t_i}{w_i}\ln10~~; \\
\slabel{eq:dRyy}
&& \frac{\partial^2R_i}{\partial y_C^2}=0~~; \\
\slabel{eq:dRgg}
&& \frac{\partial^2R_i}{\partial\gamma^2}=0~~; \\
\slabel{eq:dRbb}
&& \frac{\partial^2R_i}{\partial\beta^2}=0~~; \\
\slabel{eq:dRaa}
&& \frac{\partial^2R_i}{\partial\alpha^2}=
\frac{\beta-\gamma}\alpha\times \nonumber \\
&& \left[\frac2{\alpha^2}s_i-\frac2\alpha(x_i-
x_C)t_i+(x_i-x_C)^2\frac{t_i^2}{w_i}\ln10
\right]~~;  \\
\slabel{eq:dRyx}
&& \frac{\partial^2R_i}{\partial y_C\partial
x_C}=0~~; \\
\slabel{eq:dRyg}
&& \frac{\partial^2R_i}{\partial y_C\partial
\gamma}=0~~; \\
\slabel{eq:dRyb}
&& \frac{\partial^2R_i}{\partial y_C\partial
\beta}=0~~; \\
\slabel{eq:dRya}
&& \frac{\partial^2R_i}{\partial y_C\partial
\alpha}=0~~; \\
\slabel{eq:dRga}
&& \frac{\partial^2R_i}{\partial\gamma\partial
\alpha}=\frac1{\alpha^2}s_i-\frac{x_i-x_C}
\alpha t_i~~; \\
\slabel{eq:dRgx}
&& \frac{\partial^2R_i}{\partial\gamma\partial
x_C}=-\frac{t_i}{w_i}~~; \\
\slabel{eq:dRbx}
&& \frac{\partial^2R_i}{\partial\beta\partial
x_C}=-t_i~~; \\
\slabel{eq:dRba}
&& \frac{\partial^2R_i}{\partial\beta\partial
\alpha}=-\frac1{\alpha^2}s_i+\frac{x_i-x_C}
\alpha t_i~~; \\
\slabel{eq:dRxa}
&& \frac{\partial^2R_i}{\partial x_C\partial
\alpha}=-(\beta-\gamma)(x_i-x_C)\frac{t_i^2}
{w_i}\ln10~~;
\label{seq:dRu}
\end{leftsubeqnarray}
which allows the calculation of the Hessian
determinant.   The related, analytical
expression appears to be extremely complicated,
and for this reason it shall not be calculated
here.

In principle, the above procedure could be
used for minimizing the sum of absolute value
residuals, but at a higher price.   This is
why the first derivatives of the function:
\begin{equation}
\nonumber
F(x_C,y_C,\alpha,\beta,\gamma)=\sum_{i=1}^n
\vert R_i\vert~~;
\end{equation}
are not defined at $R_i=0$, $1\le i\le n$,
which makes a necessary condition for the
existence of an absolute minimum, expressed
by Eq.\,(\ref{eq:sid0}), satisfied at any
point, ${\sf P}^\ast(x_C^\ast,y_C^\ast,
\alpha^\ast,\beta^\ast,\gamma^\ast)$, where
$R_i=0$.   To avoid this inconvenient, our
attention has been restricted to the
minimization of the sum of square residuals.

\begin{thebibliography}{}
\bibitem{} Ascasibar, Y., Yepes, G., Gottl\"ober, S., M\"uller, V.
           2003. MRNAS 352, 1109
 \bibitem{} Bullock, J. S., Kolatt, T. S., Sigad, Y., et al.
             2001, MNRAS 321, 559
  \bibitem{} Burkert, A. 1995, ApJ 447, L25
   \bibitem{} Caimmi, R., Marmo, C. 2003, NewA 8, 119
  \bibitem{} Caimmi, R., Marmo, C., Valentinuzzi, T. 2004, astro-ph/0411455 
             (CMV)
  \bibitem{} Cole, S., Lacey, C., 1996, MNRAS 281, 716
  \bibitem{} Diemand, J., Moore B., Stadel, J. 2004, MNRAS 353, 624 
  \bibitem{} El-Zant, A.A., Hoffman, J., Primack, J., Combes, F.,
             Sloshman, I. 2004, ApJ 607, L75
  \bibitem{} Fukushige, F., Makino, J. 2001, ApJ 557, 533
  \bibitem{} Fukushige, F., Makino, J. 2003, ApJ 588, 674
  \bibitem{} Fukushige, F., Kawai, A., Makino, J. 2004, ApJ 606, 625
  \bibitem{} Hansen, S.H. 2004, MNRAS 352, L41
  \bibitem{} Hernquist, L. 1990, ApJ 356, 359
  \bibitem{} Hiotelis, N. 2003, MNRAS 344, 149
  \bibitem{} Huss, A., Jain, B., Steinmetz, M. 1999, ApJ 517, 64
  \bibitem{} Jing, Y.P., Suto, Y. 2000, ApJ 529, L69
  \bibitem{} Klypin, A., Kravtsov, A.V., Bullok, J.S., 
             Primack, J.R. 2001, ApJ 554, 903
 \bibitem{} Moore, B., Governato, F., Quinn, T., Stadel, J.,
            Lake, G. 1998, ApJ 499, L5  
 \bibitem{} Moore, B., Quinn, T., Governato, F., Stadel, J.,
            Lake, G. 1999, MNRAS 310, 1147
  \bibitem{} M\"ucket, J.P., Hoeft, M. 2003, A\&A 404, 809
 \bibitem{} Navarro, J.F., Frenk, C.S., White, S.D.M. 1995, MNRAS 275, 720
 \bibitem{} Navarro, J.F., Frenk, C.S., White, S.D.M. 1996, 
            ApJ 462, 563 
 \bibitem{} Navarro, J.F., Frenk, C.S., White, S.D.M. 1997, 
            ApJ 490, 493
 \bibitem{} Navarro, J.F., Hagashi, E., Power, C., et al. 2004, 
            MNRAS 349, 1039
 \bibitem{} Oliva, P.R., Terrasi, F. 1976. {\it Elaborazione Statistica
            dei Risultati Sperimentali}, Liguori Editore, Napoli
  \bibitem{} Rasia, E., Tormen, G., Moscardini, L. 2004, MNRAS 351, 237
  \bibitem{} Reed, D., Governato, F., Verde, L., et al. 2004, astro-ph/0312544
  \bibitem{} Ricotti, M. 2003, MNRAS 344, 1237
 \bibitem{} Rubi$\tilde{{\rm n}}$o-Martin, J.A., Rebolo, R., Carreira, P.,
            et al. 2003, MNRAS 341, 1084
 \bibitem{} Sievers, J.L., Bond, J.R., Cartwright, J.K., et al. 2003,
            ApJ 591, 599
 \bibitem{} Spergel, D.N., Verde, L., Peiris, H.V., et al. 2003, ApJSupp
            148, 175
  \bibitem{} Syer, D., White, S. D. M. 1998, MNRAS 293, 337
  \bibitem{} Smirnov, V., 1969, {\it Cours de Matematiques
             Superieures}, vol.\,I, Mir, Moscow
 \bibitem{} Tasitsiomi, A., Kravstov, A.V., Gottl\"oeber, S.,
            Klypin, A.A. 2004 ApJ 607, 125
 \bibitem{} Williams, L.L.R., Babul, A., Dalcanton, J.J. 2004,
            ApJ 604, 18 
  \bibitem{} Zhao, D.H., Mo, H.J., Jing, J.P., B\"orner, G. 2003, 
             MNRAS 339, 12

\end{thebibliography}
\end{document}